\documentclass[12pt,twoside]{article}

\usepackage{amsmath}
\usepackage{flafter}

\numberwithin{equation}{section}

\begin{document}

\title{The scalar sector of the $SU(3)_{c}\otimes SU(3)_{L}\otimes U(1)_{X}$
model}
\author{Rodolfo A. Diaz\thanks{%
radiaz@ciencias.unal.edu.co}, R. Mart\'{\i}nez\thanks{%
romart@ciencias.unal.edu.co}, and F. Ochoa\thanks{%
frecho1@hotmail.com} \and Departamento de F\'{\i}sica, Universidad Nacional, 
\\
%EndAName
Bogot\'{a}-Colombia}
\maketitle

\begin{abstract}
A complete study of the Higgs sector of the $SU(3)_{c}\otimes
SU(3)_{L}\otimes U(1)_{X}$ model is carried out, obtaining all possible
cases of vacuum expectation values that permit the spontaneous symmetry
breaking pattern$\;SU(3)_{L}\otimes U(1)_{X}\rightarrow SU(2)_{L}\otimes
U(1)_{Y}\rightarrow U(1)_{Q}.$ We find the most general Higgs potentials
that contain three triplets of Higgs and one sextet. A detailed study of the
scalar sector for different models with three Higgs triplets is done. The
models end up in an electroweak two Higgs doublet model after the first
symmetry breakdown; we find that the low energy limit depends on a trilinear
parameter of the Higgs potential, and that the decoupling limit from the
electroweak two Higgs doublet model to the minimal standard model can be
obtained quite naturally.


\end{abstract}

\section{Introduction}

Though the Standard Model (SM) \cite{sm} is a good phenomenological theory
and coincides very well with all experimental results \cite{one}, it leaves
several unanswer questions which suggest that SM should be an effective
model at low energies, originated from a more fundamental theory. Some of
the unexplained aspects in SM are: the existence of three families \cite{two}%
, \cite{three}, the mass hierarchy problem\cite{two}, \cite{three}, the
quantization of the electric charge \cite{QEC}, the large number of free
parameters to fit the model, the absence of an explanation for the matter
anti-matter asymmetry \cite{two}, the fact that the SM says nothing about
the stability of the proton, gravity cannot be incorporated as a gauge
theory \cite{three}, \cite{four}, it cannot account on the neutrino deficit
problem \cite{five}, \cite{six}, etc.

Some of these problems can be understood introducing a larger particle
content or enlarging the group of symmetry where the SM is embedded. The $%
SU(5)$\ grand unification model of Georgi and Glashow \cite{seven} unifies
the interactions and predicts the electric charge quantization; the group $%
E_{6}$ unifies the electroweak and strong interactions and might explain the
masses of the neutrinos \cite{nine}. Nevertheless, such models are for only
one fermion family. Models with larger symmetries that may allow to
understand the origin of the families have been proposed\cite{familia}. In
some models, it is also possible to understand the number of families from
the cancelation of quiral anomalies, necessary to preserve the
renormalizability of the theory\cite{anomalias}. This is the case of the $%
SU(3)_{c}\otimes SU(3)_{L}\otimes U(1)_{X},$ or 331 models, which are an
inmediate extension of the SM \cite{ten}, \cite{eleven}, \cite{twelve}, \cite%
{thirteen}. There is a great variety of such models, which have generated
new expectatives and possibilities to solve several problems of the SM.

Refs. \cite{fourteen} have studied the different 331 models based on the
criterium of cancelation of anomalies. Cancelation of chiral anomalies
demands some conditions, for instance, the number of fermionic triplets must
be equal to the number of anti-triplets\cite{doff}; notwithstanding, an
infinite number of models with exotic charges are found. However, only two
331 models with no exotic charges exist, and they are classified according
to the fermion assignment. In the present work, we find those possible 331
models from a different point of view.

As well as the cancelation of anomalies, the scalar sector should be taken
into account. This sector couples to the fermions by means of the Yukawa
Lagrangian generating additional constraints for the quantum numbers. From
the Yukawa Lagrangian, the scalar fields neccesary to endow fermions with
masses are selected. The vacuum is aligned in such a way that it respects
the symmetry breaking scheme 
\begin{equation}
SU(3)_{L}\otimes U(1)_{X}\rightarrow SU(2)_{L}\otimes U(1)_{Y}\rightarrow
U(1)_{Q}\;\;(31\rightarrow 21\rightarrow 1)
\end{equation}%
where the different solutions to align the vacuum fit quantum numbers which
are important in the determination of the electric charge and to determine
the different 331 models.

In section \ref{FV repres}, we examine the structure of the representations
for fermions and vector bosons under the quiral group $SU(3)_{c}\otimes $ $%
SU(3)_{L}\otimes U(1)_{X}$.\ In section \ref{S repres}, we analize the
representations in the Higgs sector, as well as the possible vacuum
alignments that generates the scheme of spontaneous symmetry breaking (SSB) $%
31\rightarrow 21\rightarrow 1$. Along with these possibilities, section \ref%
{Higgs pot} shows the most general Higgs potentials for any type of 331
models containing three Higgs triplets and one Higgs sextet, necessary for
all the fermions to acquire their masses. In section \ref{S masses}, we
apply our general results about the scalar sector for a model of type $\beta
=1/\sqrt{3}$ with three Higgs triplets, obtaining the complete scalar
spectrum. Section \ref{low energy} shows the behavior of the $\beta =1/\sqrt{%
3}$ model in the low energy regime. Finally, section \ref{conclusions} is
regarded for our conclusions.

\section{Fermion and vector representations\label{FV repres}}

\subsection{Fermion representations}

The fermions present the following general structure of transformations
under the quiral group $SU(3)_{c}\otimes $ $SU(3)_{L}\otimes U(1)_{X}$

\begin{eqnarray}
\widehat{\psi }_{L} &=&\left\{ 
\begin{array}{c}
\widehat{q}_{L}:\left( \mathbf{3,3,}X_{q}^{L}\right) =\left( \mathbf{3,2,}%
X_{q}^{L}\right) \oplus \left( \mathbf{3,1,}X_{q}^{L}\right) , \\ 
\widehat{\ell }_{L}:\left( \mathbf{1,3,}X_{\ell }^{L}\right) =\left( \mathbf{%
1,2,}X_{\ell }^{L}\right) \oplus \left( \mathbf{1,1,}X_{\ell }^{L}\right) ,%
\end{array}%
\right.  \notag \\
\widehat{\psi }_{L}^{\ast } &=&\left\{ 
\begin{array}{c}
\widehat{q}_{L}^{\ast }:\left( \mathbf{3,3}^{\ast }\mathbf{,-}%
X_{q}^{L}\right) =\left( \mathbf{3,2}^{\ast }\mathbf{,-}X_{q}^{L}\right)
\oplus \left( \mathbf{3,1,-}X_{q}^{L}\right) \mathbf{,} \\ 
\widehat{\ell }_{L}^{\ast }:\left( \mathbf{1,3}^{\ast }\mathbf{,-}X_{\ell
}^{L}\right) =\left( \mathbf{1,2}^{\ast }\mathbf{,-}X_{\ell }^{L}\right)
\oplus \left( \mathbf{1,1,-}X_{\ell }^{L}\right) ,%
\end{array}%
\right.  \notag \\
\widehat{\psi }_{R} &=&\left\{ 
\begin{array}{c}
\widehat{q}_{R}:\left( \mathbf{3,1,}X_{q}^{R}\right) , \\ 
\widehat{\ell }_{R}:\left( \mathbf{1,1,}X_{\ell }^{R}\right) ,%
\end{array}%
\right.  \label{1}
\end{eqnarray}%
where the quarks $\widehat{q}$ are color triplets (\textbf{3}) and the
leptons $\widehat{\ell }$ are color singlets (\textbf{1}). The second
equality corresponds to the branching rules $SU(2)_{L}\subset SU(3)_{L}$.
Both possibilities \textbf{3 }and \textbf{3}$^{\ast }$ are included in the
flavor sector $(SU(3)_{L})$, since the same number of fermion triplets and
antitriplets must be present in order to cancel anomalies \cite{doff}. The
generator of $U(1)_{X}$ conmute with the matrices of $SU(3)_{L}$, therefore
it should be a matrix proportional to the identity, where the factor of
proporcionality $X_{p}$ takes a value according to the representations of $%
SU(3)_{L}$ and the anomalies cancelation.

The electric charge is defined in general as a linear combination of the
diagonal generators of the group 
\begin{equation}
\widehat{Q}=a\widehat{T}_{3}+\beta \widehat{T}_{8}+X\widehat{I},  \label{2}
\end{equation}%
with $T_{3}=\frac{1}{2}diag(1,-1,0)$ and $T_{8}=\frac{1}{2\sqrt{3}}%
diag(1,1,-2),$ where the normalization chosen is $Tr(T_{\alpha }T_{\beta })=%
\frac{1}{2}\delta _{\alpha \beta },$ and $I=diag\left( 1,1,1\right) $ is the
identity matrix. In this work we shall show that these quantum numbers can
also be fixed by choicing the scalar sector and vacuum alignment to break
the symmetry of the model.

\subsection{Representation of gauge bosons}

The gauge bosons associated to the group $SU(3)_{L}$ transform according to
the adjoint representation and are written in the form 
\begin{equation}
\mathbf{W}_{\mu }=W_{\mu }^{\alpha }G_{\alpha }=\frac{1}{2}\left[ 
\begin{array}{ccc}
W_{\mu }^{3}+\frac{1}{\sqrt{3}}W_{\mu }^{8} & \sqrt{2}W_{\mu }^{+} & \sqrt{2}%
K_{\mu }^{Q_{1}} \\ 
\sqrt{2}W_{\mu }^{-} & -W_{\mu }^{3}+\frac{1}{\sqrt{3}}W_{\mu }^{8} & \sqrt{2%
}K_{\mu }^{Q_{2}} \\ 
\sqrt{2}K_{\mu }^{-Q_{1}} & \sqrt{2}K_{\mu }^{-Q_{2}} & -\frac{2}{\sqrt{3}}%
W_{\mu }^{8}%
\end{array}%
\right] .  \label{3}
\end{equation}%
Therefore, the electric charge takes the general form 
\begin{equation}
Q_{W}\rightarrow \left[ 
\begin{array}{ccc}
0 & 1 & \frac{1}{2}+\frac{\sqrt{3}\beta }{2} \\ 
-1 & 0 & -\frac{1}{2}+\frac{\sqrt{3}\beta }{2} \\ 
-\frac{1}{2}-\frac{\sqrt{3}\beta }{2} & \frac{1}{2}-\frac{\sqrt{3}\beta }{2}
& 0%
\end{array}%
\right] .  \label{4}
\end{equation}%
From Eq. (\ref{4}) we see that the cases $\beta =\pm \sqrt{3}$ lead to
exotic charges and only the cases $\beta =\pm 1/\sqrt{3}$ does not. As for
the gauge field associated to $U(1)_{X},$ it is represented as 
\begin{equation}
\mathbf{B}_{\mu }=\mathbf{I}_{3\times 3}B_{\mu },  \label{5}
\end{equation}%
which is a singlet under $SU(3)_{L},$ and its electric charge is given by 
\begin{equation}
Q_{B_{\mu }}\rightarrow 0.  \label{6}
\end{equation}%
From the previous expressions we see that three gauge fields with charges
equal to zero are obtained, and in the basis of mass eigenstates they
correspond to the photon, $Z$ and $Z^{\prime }$. Two fields with charges $%
\pm 1$ associated to $W^{\pm }$ and four fields with charges that depend on
the choice of $\beta $. Demanding that the model contains no exotic charges
is equivalent to settle $\beta =-1/\sqrt{3}$ \cite{twelve}, and $\beta =1/%
\sqrt{3}$. The Higgs potential of the model with $\beta =1/\sqrt{3}$ has
been discussed recently by the third of Refs. \cite{fourteen}, however we
consider the most general Higgs potential with three Higgs triplets and we
also show that it reduces to the 2HDM in a natural way when the decoupling
limit is taken.

In both of the cases $\beta =\pm 1/\sqrt{3}$, the fields $K_{\mu }^{\pm
Q_{1}}$ and $K_{\mu }^{\pm Q_{2}}$ posess charges $\pm 1$ and $0$ . It is
important to take into account the scalar sector and the symmetry breakings
to fix this quantum number, which in turn determine the would-be Goldstone
bosons associated to the gauge fields, with the same electric charge of the
gauge fields that are acquiring mass in the different scales of breakdown.
When $\beta $ takes a value different from the previous ones we get fields
with exotic electric charges.

\subsection{Fermionic anomalies}

The coefficient of triangular anomalies is defined as

\begin{eqnarray}
\frac{1}{2}A_{\alpha \beta \gamma } &=&Tr\left[ \left\{ G_{\alpha }(\widehat{%
T}_{\psi })_{L},G_{\beta }(\widehat{T}_{\psi })_{L}\right\} G_{\gamma }(%
\widehat{T}_{\psi })_{L}\right.  \notag \\
&&\left. -\left\{ G_{\alpha }(\widehat{T}_{\psi })_{R},G_{\beta }(\widehat{T}%
_{\psi })_{R}\right\} G_{\gamma }(\widehat{T}_{\psi })_{R}\right]
\label{anomalcond}
\end{eqnarray}%
where $G_{\alpha }\left( \widehat{T}_{\psi }\right) _{h}$ are the matrix
representations for each group generator acting on the basis $\psi $ with
helicity left or right. In the particular case of the $331$ gauge models,
such conditions become%
\begin{equation}
\left. 
\begin{array}{l}
(a)\qquad \left[ SU(3)_{c}\right] ^{3}\mapsto A_{\alpha \beta \gamma }=0, \\ 
(b)\qquad \left[ SU(3)_{c}\right] ^{2}\otimes SU(3)_{L}\mapsto A_{\alpha
\beta \gamma }=0, \\ 
(c)\qquad \left[ SU(3)_{c}\right] ^{2}\otimes U(1)_{X}\mapsto \frac{1}{2}%
A_{\alpha \alpha 0}=\pm 3X_{q}^{L}-\sum_{\text{sing}}X_{q}^{R}=0, \\ 
(d)\qquad SU(3)_{c}\otimes \left[ SU(3)_{L}\right] ^{2}\mapsto A_{\alpha
\beta \gamma }=0, \\ 
(e)\qquad SU(3)_{c}\otimes \left[ U(1)_{X}\right] ^{2}\mapsto A_{00\gamma
}=0, \\ 
(f)\qquad SU(3)_{c}\otimes SU(3)_{L}\otimes U(1)_{X}\mapsto \frac{1}{2}%
A_{\alpha \alpha 0}=0, \\ 
(g)\qquad \left[ SU(3)_{L}\right] ^{3}\mapsto \frac{1}{2}A_{\alpha \beta
\gamma }=0, \\ 
(h)\qquad \left[ SU(3)_{L}\right] ^{2}\otimes U(1)_{X}\mapsto \frac{1}{2}%
A_{\alpha \alpha 0}=\sum\limits_{m}\left( \pm X_{\ell ^{(m)}}^{L}\right) \pm
3X_{q}^{L}=0, \\ 
(i)\qquad SU(3)_{L}\otimes \left[ U(1)_{X}\right] ^{2}\mapsto A_{00\gamma
}=0, \\ 
(j)\qquad \left[ U(1)_{X}\right] ^{3}\mapsto \frac{1}{4}A_{000}=3\sum%
\limits_{m}\left( \pm X_{\ell }^{L}\right) ^{3}\pm 9\left( X_{q}^{L}\right)
^{3} \\ 
\qquad \qquad \qquad \qquad \qquad \qquad \ \ -\sum_{\text{sing}}\left(
X_{\ell }^{R}\right) ^{3}-3\sum_{\text{sing}}\left( X_{q}^{R}\right) ^{3}=0.%
\end{array}%
\right.  \label{anomalcond2}
\end{equation}%
When we impose for the fermionic representations of $SU\left( 3\right)
_{c}\otimes SU\left( 3\right) _{L}\ $to be vector-like, then the number of
color triplets should be equal to the number of color anti-triplets, and
same for the left triplets. In this case, all the conditions of Eqs. (\ref%
{anomalcond2}) are satisfied, except the ones that involve the quantum
numbers asociated to $U\left( 1\right) _{X}$. Considering a particle content
as the one in table \ref{tab:fermionnum}; the non-trivial conditions are
reduced to%
\begin{align*}
\frac{1}{2}A_{\alpha \alpha 0}& =\pm 3X_{q}^{L}-\sum_{\text{singlets}%
}X_{q}^{R}=0 \\
\frac{1}{2}A_{\alpha \alpha 0}& =\sum\limits_{m}\left( \pm X_{\ell
^{(m)}}^{L}\right) \pm 3X_{q}^{L}=0 \\
\frac{1}{4}A_{000}& =3\sum\limits_{m}\left( \pm X_{\ell ^{(m)}}^{L}\right)
^{3}\pm 9\left( X_{q}^{L}\right) ^{3} \\
& -\sum_{\text{singlets}}\left( X_{\ell }^{R}\right) ^{3}-3\sum_{\text{%
singlets}}\left( X_{q}^{R}\right) ^{3}=0 \\
\left[ Grav\right] ^{2}\otimes U(1)_{X}& \mapsto
A_{grav}=3\sum\limits_{m}\left( \pm X_{\ell ^{(m)}}^{L}\right) \pm 9X_{q}^{L}
\\
& -\sum_{\text{singlets}}\left( X_{\ell }^{R}\right) -3\sum_{\text{singlets}%
}\left( X_{q}^{R}\right) =0
\end{align*}%
where we have included the gravitational anomaly condition as well. These
conditions still permit an infinite number of solutions for the quantum
numbers, that can be characterized by the values of $a$ and $\beta $ in the
definition of the electromagnetic charge Eq. (\ref{2}). However, $a=1$ is
required to obtain isospin doublets to embed properly the $SU\left( 2\right)
_{L}\otimes U\left( 1\right) _{Y}$ model into $SU\left( 3\right) _{L}\otimes
U\left( 1\right) _{X}$. Now, in order to restrict $\beta $, we could demand
that the exotic component of the triplet has ordinary electromagnetic
charge, in whose case $\beta =\pm 1/\sqrt{3}$. On the other hand, we can
obtain relations between $\beta $ and the quantum numbers associated to $%
U\left( 1\right) _{X}$ by imposing gauge invariance in the Yukawa sector,
and by taking into account the symmetry breaking pattern.

\begin{center}
\begin{table}[!tbh]
\begin{center}
$%
\begin{tabular}{||c||c||c||}
\hline\hline
& $Q_{\psi }$ & $Y_{\psi }$ \\ \hline\hline
$q_{L}=\left( 
\begin{array}{c}
q^{1} \\ 
q^{2} \\ 
q^{3}%
\end{array}%
\right) _{L}=\left( 
\begin{array}{c}
q^{\delta } \\ 
q^{3}%
\end{array}%
\right) _{L}$ & $\left( 
\begin{array}{c}
\pm \frac{2}{3}\pm X_{q}^{L} \\ 
\mp \frac{1}{3}\pm X_{q}^{L} \\ 
\mp \frac{1}{3}\pm X_{q}^{L}%
\end{array}%
\right) $ & $\left( 
\begin{array}{c}
\pm \frac{1}{6}\pm X_{q}^{L} \\ 
\pm \frac{1}{6}\pm X_{q}^{L} \\ 
\mp \frac{1}{3}\pm X_{q}^{L}%
\end{array}%
\right) $ \\ \hline\hline
$\ell _{L}^{(m)}=\left( 
\begin{array}{c}
\ell ^{1(m)} \\ 
\ell ^{2(m)} \\ 
\ell ^{3(m)}%
\end{array}%
\right) _{L}=\left( 
\begin{array}{c}
\ell ^{\delta (m)} \\ 
\ell ^{3(m)}%
\end{array}%
\right) _{L}$ & $\left( 
\begin{array}{c}
\pm \frac{2}{3}\pm X_{\ell ^{(m)}}^{L} \\ 
\mp \frac{1}{3}\pm X_{\ell ^{(m)}}^{L} \\ 
\mp \frac{1}{3}\pm X_{\ell ^{(m)}}^{L}%
\end{array}%
\right) $ & $\left( 
\begin{array}{c}
\pm \frac{1}{6}\pm X_{\ell ^{(m)}}^{L} \\ 
\pm \frac{1}{6}\pm X_{\ell ^{(m)}}^{L} \\ 
\mp \frac{1}{3}\pm X_{\ell ^{(m)}}^{L}%
\end{array}%
\right) $ \\ \hline\hline
$q_{R}\rightarrow \left\{ 
\begin{array}{c}
q_{R}^{1} \\ 
q_{R}^{2} \\ 
q_{R}^{3}%
\end{array}%
\right. $ & $\pm X_{q}^{R}$ & $\pm X_{q}^{R}$ \\ \hline\hline
$\ell _{R}^{(n)}$ & $X_{\ell _{n}}^{R}$ & $X_{\ell _{n}}^{R}$ \\ \hline\hline
\end{tabular}%
$%
\end{center}
\caption{\textit{Particle content for the fermionic sector of the 331 model
for a one family scenario. $q$ refers to triplets of color, $m=1,2,3$ and $n$
is arbitrary.}}
\label{tab:fermionnum}
\end{table}
\end{center}

\section{Scalar representations\label{S repres}}

\subsection{SSB scheme}

From the phenomenological point of view, the SM is an effective theory at
the electroweak scale, hence, the SSB follows the scheme 
\begin{equation}
SU(3)_{L}\otimes U(1)_{X}\overset{\Phi _{1}}{\longrightarrow }%
SU(2)_{L}\otimes U(1)_{Y}\overset{\Phi _{2}}{\longrightarrow }U(1)_{Q}
\label{res}
\end{equation}%
In the first breakdown there are five gauge fields that acquire mass
proportional to $<\Phi _{1}>$ and in the second breakdown (electroweak
breakdown), three gauge fields acquire mass of the order of $<\Phi _{2}>$,
leaving a massless field associated to the unbroken generator $Q$, the
photon.

\subsubsection{$SU(3)_{L}\otimes U(1)_{X}\rightarrow SU(2)_{L}\otimes
U(1)_{Y}\;$transition}

In the first transition the VEV's of $\Phi _{1}$ break the symmetry $%
SU(3)_{L}\otimes U(1)_{X}/SU(2)_{L}\otimes U(1)_{Y}$ where the hypercharge
is defined as $\widehat{Y}=\beta \widehat{T}_{8}+X\widehat{I}$. The
conditions that should be satisfied in this breaking step are 
\begin{equation}
\begin{array}{l}
\left[ \widehat{T}_{1}^{L},\left\langle \Phi _{1}\right\rangle _{0}\right] =0
\\[3mm] 
\left[ \widehat{T}_{2}^{L},\left\langle \Phi _{1}\right\rangle _{0}\right] =0
\\[3mm] 
\left[ \widehat{T}_{3}^{L},\left\langle \Phi _{1}\right\rangle _{0}\right] =0
\\[3mm] 
\left[ \beta \widehat{T}_{8}^{L}+X\widehat{I},\left\langle \Phi
_{1}\right\rangle _{0}\right] =0%
\end{array}%
,%
\begin{array}{l}
\left[ \widehat{T}_{4}^{L},\left\langle \Phi _{1}\right\rangle _{0}\right]
\neq 0 \\[3mm] 
\left[ \widehat{T}_{5}^{L},\left\langle \Phi _{1}\right\rangle _{0}\right]
\neq 0 \\[3mm] 
\left[ \widehat{T}_{6}^{L},\left\langle \Phi _{1}\right\rangle _{0}\right]
\neq 0 \\[3mm] 
\left[ \widehat{T}_{7}^{L},\left\langle \Phi _{1}\right\rangle _{0}\right]
\neq 0 \\[3mm] 
\left[ \beta \widehat{T}_{8}^{L}-X\widehat{I},\left\langle \Phi
_{1}\right\rangle _{0}\right] \neq 0%
\end{array}
\label{primerromp}
\end{equation}%
These conditions reflect the fact that five bosons acquire mass and in order
to preserve the number of degrees of freedom, at least five components in $%
\Phi _{1}$ are needed to represent the would be Goldstone bosons.

\subsubsection{$SU(2)_{L}\otimes U(1)_{Y}\rightarrow U(1)_{Q}\;$transition}

In the second transition the VEV's of $\Phi _{2}$ breaks the symmetry $%
SU(2)_{L}\otimes U(1)_{Y}/U(1)_{Q}$. The conditions that should be satisfied
in this breakdown read 
\begin{equation}
\left[ Q,\left\langle \Phi _{2}\right\rangle _{0}\right] =0, 
\begin{array}{l}
\left[ \widehat{T}_{1}^{L},\left\langle \Phi _{2}\right\rangle _{0}\right]
\neq 0 \\[3mm] 
\left[ \widehat{T}_{2}^{L},\left\langle \Phi _{2}\right\rangle _{0}\right]
\neq 0 \\[3mm] 
\left[ \widehat{T}_{3}^{L}-(\beta \widehat{T}_{8}^{L}+X\widehat{I}%
),\left\langle \Phi _{2}\right\rangle _{0}\right] \neq 0%
\end{array}
,  \label{segundoromp}
\end{equation}
where three gauge bosons acquire mass, from which $\Phi _{2}$ needs three
components associated to the would be Goldstone bosons.

\subsection{Yukawa terms}

Other conditions fixed by the scalar sector are related to the
fermion-fermion-scalar couplings that generate the fermion masses. The
scalar fields have to be coupled to fermions by Yukawa terms invariant under 
$SU(3)_{L}\otimes U(1)_{X}$; these couplings can be written as 
\begin{eqnarray}
\overline{\psi _{L}^{i}}\psi _{R}\Phi &:&3^{\ast }\otimes 1\otimes \Phi
=1\Rightarrow \;\;\Phi =3,  \notag \\
\overline{\psi _{L}^{i}}\left( \psi _{\,L}^{j}\right) ^{c}\Phi &:&3^{\ast
}\otimes 3^{\ast }\otimes \Phi =1\Rightarrow \;\;\Phi =3\otimes 3=3^{\ast
}\oplus 6,  \notag \\
\overline{\psi _{R}}\left( \psi _{R}\right) ^{c}\Phi &:&1\otimes 1\otimes
\Phi =1\Rightarrow \Phi =1,  \notag \\
\overline{\left( \psi _{R}\right) ^{c}}\left( \psi _{L}^{i}\right) ^{c}\Phi
&:&1\otimes 3^{\ast }\otimes \Phi =1\Rightarrow \Phi =3.  \label{yukawas}
\end{eqnarray}%
Hence, in order to generate the masses of the fermions, the Higgs bosons
should lie either in the singlet, triplet, antitriplet or sextet
representations of $SU(3)_{L}$.

\subsection{Scalar representations for the first transition ($\Phi _{1}$)}

The choice of the scalar sector should fulfill the following basic conditions

\begin{enumerate}
\item The VEV's should accomplish the conditions (\ref{primerromp}) and (\ref%
{segundoromp}) in the first and second transition respectively.

\item When the VEV's are replaced in Eq. (\ref{yukawas}), the fermions
should acquire the appropiate masses.

\item The number of components of $\Phi $ should be at least the number of
would be Goldstone bosons for each transition of SSB.
\end{enumerate}

\subsubsection{Triplet representation \textbf{3\label{triplet 3}}}

One of the possible solutions of the Eq. (\ref{yukawas}) is the fundamental
representation \textbf{3} where the VEV's can be written in the form 
\begin{equation}
\left\langle \chi \right\rangle _{0}=\left( 
\begin{array}{c}
\nu _{\chi _{1}} \\ 
\nu _{\chi _{2}} \\ 
\nu _{\chi _{3}}%
\end{array}%
\right) .
\end{equation}%
By demanding the conditions of Eq. (\ref{primerromp}) for the first breaking
of symmetry, it is found that the vacuum should be aligned in the following
way 
\begin{equation}
\left\langle \chi \right\rangle _{0}=\left( 
\begin{array}{c}
0 \\ 
0 \\ 
\nu _{\chi _{3}}%
\end{array}%
\right)  \label{triplete1}
\end{equation}%
with the condition 
\begin{equation}
\quad X_{\chi }-\frac{\beta }{\sqrt{3}}=0,\quad \beta \neq 0\quad \nu _{\chi
_{3}}\neq 0,  \label{triplewtecon1}
\end{equation}%
The choice of vacuum above generates the following mass terms 
\begin{eqnarray}
\mathcal{L}_{Y} &=&\Gamma _{1}^{\chi }\overline{\psi _{L}^{i}}\psi
_{R}\langle \chi ^{i}\rangle +\Gamma _{2}^{\chi }\overline{\left( \psi
_{R}\right) ^{c}}\left( \psi _{L}^{i}\right) ^{c}\langle \chi ^{i}\rangle
+h.c. \\
&=&\Gamma _{1}^{\chi }\nu _{\chi ^{3}}\overline{\psi _{L}^{3}}\psi
_{R}+\Gamma _{2}^{\chi }\nu _{\chi ^{3}}\overline{\left( \psi _{R}\right)
^{c}}\left( \psi _{L}^{3}\right) ^{c}+h.c.  \label{17}
\end{eqnarray}

\subsubsection{Antisymmetric representation \textbf{3}$^{\ast }$}

Other possible representation to generate masses from Eq. (\ref{yukawas}) is
the antisymmetric one \textbf{3}$^{\ast }$. If we use this representation
for the first breaking of the symmetry, then the conditions of Eq. (\ref%
{primerromp}) should be satisfied, implying for the vacuum to be aligned as 
\begin{equation}
\left\langle \chi ^{\ast }\right\rangle _{0}=\left( 
\begin{array}{c}
0 \\ 
0 \\ 
\nu _{\chi _{3^{\ast }}}%
\end{array}%
\right)
\end{equation}%
with the condition 
\begin{equation}
\quad X_{\chi ^{\ast }}-\frac{\beta }{\sqrt{3}}=0,\quad \beta \neq 0;\quad
\nu _{\chi _{3^{\ast }}}\neq 0.  \label{12}
\end{equation}%
The anti-triplet antisymmetric representation only admits mass Yukawa terms
of the form 
\begin{eqnarray}
\mathcal{L}_{Y} &=&\Gamma ^{\Phi _{1}}\epsilon _{ijk}\overline{\psi _{L}^{i}}%
\left( \psi _{L}^{j}\right) ^{c}\langle \chi ^{k}\rangle +h.c  \notag \\
&=&\Gamma ^{\chi }\nu _{\chi _{3^{\ast }}}\epsilon _{ij3}\overline{\psi
_{L}^{i}}\left( \psi _{L}^{j}\right) ^{c}+h.c.  \label{13}
\end{eqnarray}

\subsubsection{Symmetric representation \textbf{6}}

As for the possibility of $\Phi _{1}=S$ symmetric, we have:

\begin{equation}
\left\langle S^{ij}\right\rangle _{0}=\left[ 
\begin{array}{ccc}
\nu _{1} & \nu _{2} & \nu _{3} \\ 
\nu _{2} & \nu _{4} & \nu _{5} \\ 
\nu _{3} & \nu _{5} & \nu _{6}%
\end{array}%
\right] ,
\end{equation}%
and requiring the conditions of Eq. (\ref{primerromp}), it is obtained 
\begin{equation}
\left\langle S^{ij}\right\rangle _{0}=\left[ 
\begin{array}{ccc}
0 & 0 & 0 \\ 
0 & 0 & 0 \\ 
0 & 0 & \nu _{6}%
\end{array}%
\right]
\end{equation}%
with 
\begin{equation}
\quad X_{S^{ij}}-\frac{\beta }{\sqrt{3}}=0,\quad \beta \neq 0\quad \nu
_{6}\neq 0.  \label{14}
\end{equation}%
The mass terms generated from the Yukawa Lagrangian are given by 
\begin{equation}
\mathcal{L}_{Y}=\Gamma ^{\Phi _{1}}\overline{\psi _{L}^{i}}\left( \psi
_{L}^{j}\right) ^{c}\langle S^{ij}\rangle +h.c=\Gamma ^{\chi }\nu _{6}%
\overline{\psi _{L}^{3}}\left( \psi _{L}^{3}\right) ^{c}+h.c.  \label{15}
\end{equation}

\subsection{Scalar representations for the second transition ($\Phi _{2}$)}

The first breaking align the vacuum according to the conditions (\ref%
{primerromp}) and impose a requirement on the quantum numbers of the
representations $X[\Phi ]-\beta /\sqrt{3}=0$. This condition gives freedom
to choice the electric charge of the fields of the scalar representations in
different directions. To align the vacuum in the second breakdown according
to the conditions (\ref{segundoromp}) we shall write explicitly the electric
charge of the components of the scalar fields. 
\begin{equation*}
\left[ \widehat{Q},\Phi \right] =\left[ \widehat{T}_{3},\Phi \right] +\beta %
\left[ \widehat{T}_{8},\Phi \right] +\left[ \widehat{X},\Phi \right] .
\end{equation*}%
The charges for the representations $3$, $3^{\ast }$ and $6$ are given
respectively by 
\begin{eqnarray}
Q_{3} &=&\left( 
\begin{array}{c}
\frac{1}{2}+\frac{\beta }{2\sqrt{3}}+X_{\Phi } \\ 
-\frac{1}{2}+\frac{\beta }{2\sqrt{3}}+X_{\Phi } \\ 
-\frac{\beta }{\sqrt{3}}+X_{\Phi }%
\end{array}%
\right) ,  \label{carga3} \\
Q_{3^{\ast }} &=&\left( 
\begin{array}{c}
-\frac{1}{2}-\frac{\beta }{2\sqrt{3}}-X_{\Phi ^{\ast }} \\ 
\frac{1}{2}-\frac{\beta }{2\sqrt{3}}-X_{\Phi ^{\ast }} \\ 
\frac{\beta }{\sqrt{3}}-X_{\Phi ^{\ast }}%
\end{array}%
\right) ,  \label{carga3*} \\
Q_{6} &=&\left[ 
\begin{array}{ccc}
1+\frac{\beta }{\sqrt{3}}+2X_{\Phi ^{ij}} & \frac{\beta }{\sqrt{3}}+2X_{\Phi
^{ij}} & \frac{1}{2}-\frac{\beta }{2\sqrt{3}}+2X_{\Phi ^{ij}} \\ 
\frac{\beta }{\sqrt{3}}+2X_{\Phi ^{ij}} & -1+\frac{\beta }{\sqrt{3}}%
+2X_{\Phi ^{ij}} & -\frac{1}{2}-\frac{\beta }{2\sqrt{3}}+2X_{\Phi ^{ij}} \\ 
\frac{1}{2}-\frac{\beta }{2\sqrt{3}}+2X_{\Phi ^{ij}} & -\frac{1}{2}-\frac{%
\beta }{2\sqrt{3}}+2X_{\Phi ^{ij}} & -\frac{2\beta }{\sqrt{3}}+2X_{\Phi
^{ij}}%
\end{array}%
\right] .  \label{carga6}
\end{eqnarray}%
We should notice that the choice of a particular vacuum alignment fixes the
value of the quantum number $X_{\Phi },$ and it in turn fixes the $\beta $
parameter according to the solutions (\ref{triplewtecon1}), (\ref{12}) and (%
\ref{14}) respectively. The quantum numbers $X$ for the fermions can be
determined by means of the Yukawa sector, according to the couplings of
fermions to the scalar sector and same for the gauge fields which acquire
mass through the covariant derivative of the scalar fields. Therefore, it
should exist a relation amongst both the fermion and gauge sectors with the
scalar one. Owing to it, the quantum numbers of the scalar sector were
settled free for the different symmetry breakings to fix them and find the
different 331 models.

For this transition the vacuum should be chosen in the direction in which
the scalar fields have null charge. From Eqs. (\ref{carga3})$-$(\ref{carga6}%
) we observe that the components depend on the value assigned to $\beta $
and $X_{\Phi }$. We will align the vacuum arbitrarily and choice the quantum
numbers $\beta $ and $X_{\Phi }$ that define the electric charge operator
according to the conditions (\ref{segundoromp}).

\subsubsection{Fundamental representation 3}

Like in section \ref{triplet 3}, the possibility for a triplet is considered
but for the second transition. Thus, the conditions (\ref{segundoromp})
should be fulfilled. In table \ref{tab:uno} the possible vacuum alignments
for a representation \textbf{3} are shown.

\renewcommand{\arraystretch}{1.2}

\begin{center}
\begin{table}[!tbh]
$%
\begin{tabular}{|c|c|c|c|}
\hline
& $\beta =\frac{1}{\sqrt{3}}$ & $\beta =-\frac{1}{\sqrt{3}}$ & $\beta \neq
\pm \frac{1}{\sqrt{3}}$ \\ \hline
$\left\langle \rho \right\rangle _{0}$ & 
\begin{tabular}{cc}
$\left( 
\begin{array}{c}
0 \\ 
\nu _{\rho _{2}} \\ 
\nu _{\rho _{3}}%
\end{array}%
\right) $ & $\left( 
\begin{array}{c}
\nu _{\rho _{1}} \\ 
0 \\ 
0%
\end{array}%
\right) $%
\end{tabular}
& 
\begin{tabular}{cc}
$\left( 
\begin{array}{c}
0 \\ 
\nu _{\rho _{2}} \\ 
0%
\end{array}%
\right) $ & $\left( 
\begin{array}{c}
\nu _{\rho _{1}} \\ 
0 \\ 
\nu _{\rho _{3}}%
\end{array}%
\right) $%
\end{tabular}
& $\left( 
\begin{array}{c}
0 \\ 
\nu _{\rho _{2}} \\ 
0%
\end{array}%
\right) \quad \quad \left( 
\begin{array}{c}
\nu _{\rho _{1}} \\ 
0 \\ 
0%
\end{array}%
\right) $ \\ \hline
$X_{\rho }$ & 
\begin{tabular}{cc}
$\frac{1}{3}\quad \quad $ & $-\frac{2}{3}$%
\end{tabular}
& 
\begin{tabular}{cc}
$\frac{2}{3}\quad \quad $ & $\quad -\frac{1}{3}$%
\end{tabular}
& $\frac{1}{2}-\frac{\beta }{2\sqrt{3}}\quad \quad -\frac{1}{2}-\frac{\beta 
}{2\sqrt{3}}$ \\ \hline
\end{tabular}%
\ $%
\caption{\textit{Solutions for the second SSB with Higgs triplets in the
fundamental representation 3.}}
\label{tab:uno}
\end{table}
\end{center}

\subsubsection{Antisymmetric representation $3^{\ast }$}

For the anti-triplet antisymmetric representation, we get the same type of
solutions as in table \ref{tab:uno}, and they are shown in table \ref%
{tab:dos}.

\begin{table}[!tbh]
\begin{center}
$%
\begin{tabular}{|c|c|c|c|}
\hline
& $\beta =\frac{1}{\sqrt{3}}$ & $\beta =-\frac{1}{\sqrt{3}}$ & $\beta \neq
\pm \frac{1}{\sqrt{3}}$ \\ \hline
$\left\langle \rho ^{\ast }\right\rangle _{0}$ & 
\begin{tabular}{cc}
$\left( 
\begin{array}{c}
0 \\ 
\nu _{\rho _{2}^{\ast }} \\ 
\nu _{\rho _{3}^{\ast }}%
\end{array}%
\right) $ & $\left( 
\begin{array}{c}
\nu _{\rho _{1}^{\ast }} \\ 
0 \\ 
0%
\end{array}%
\right) $%
\end{tabular}
& 
\begin{tabular}{cc}
$\left( 
\begin{array}{c}
0 \\ 
\nu _{\rho _{2}^{\ast }} \\ 
0%
\end{array}%
\right) $ & $\left( 
\begin{array}{c}
\nu _{\rho _{1}^{\ast }} \\ 
0 \\ 
\nu _{\rho _{3}^{\ast }}%
\end{array}%
\right) $%
\end{tabular}
& $\left( 
\begin{array}{c}
0 \\ 
\nu _{\rho _{2}^{\ast }} \\ 
0%
\end{array}%
\right) \quad \quad \left( 
\begin{array}{c}
\nu _{\rho _{1}^{\ast }} \\ 
0 \\ 
0%
\end{array}%
\right) $ \\ \hline
$X_{\rho ^{\ast }}$ & 
\begin{tabular}{cc}
$\frac{1}{3}\quad \quad $ & $-\frac{2}{3}$%
\end{tabular}
& 
\begin{tabular}{cc}
$\frac{2}{3}\quad \quad $ & $\quad -\frac{1}{3}$%
\end{tabular}
& $\frac{1}{2}-\frac{\beta }{2\sqrt{3}}\quad \quad -\frac{1}{2}-\frac{\beta 
}{2\sqrt{3}}$ \\ \hline
\end{tabular}%
\ $%
\end{center}
\caption{\textit{Solutions for the second SSB with Higgs triplets in the
antisymmetric representation 3$^{\ast }$.}}
\label{tab:dos}
\end{table}

\subsubsection{Symmetric representation 6}

For the representation of dimension $6$ we have the solutions to the
conditions (\ref{segundoromp}) shown in table \ref{tab:tres}. 
\begin{equation*}
\begin{tabular}{|c|c|c|}
\hline
& $\beta =\frac{1}{\sqrt{3}}$ & $\beta =-\frac{1}{\sqrt{3}}$ \\ \hline
$\left\langle \rho ^{ij}\right\rangle _{0}$ & 
\begin{tabular}{c}
$\left( 
\begin{array}{ccc}
0 & \nu _{2} & \nu _{3} \\ 
\nu _{2} & 0 & 0 \\ 
\nu _{3} & 0 & 0%
\end{array}
\right) \left( 
\begin{array}{ccc}
0 & 0 & 0 \\ 
0 & \nu _{4} & \nu _{5} \\ 
0 & \nu _{5} & \nu _{6}%
\end{array}
\right) $ \\ 
$\left( 
\begin{array}{ccc}
\nu _{1} & 0 & 0 \\ 
0 & 0 & 0 \\ 
0 & 0 & 0%
\end{array}
\right) $%
\end{tabular}
& 
\begin{tabular}{c}
$\left( 
\begin{array}{ccc}
\nu _{1} & 0 & \nu _{3} \\ 
0 & 0 & 0 \\ 
\nu _{3} & 0 & \nu _{6}%
\end{array}
\right) \left( 
\begin{array}{ccc}
0 & \nu _{2} & 0 \\ 
\nu _{2} & 0 & \nu _{5} \\ 
0 & \nu _{5} & 0%
\end{array}
\right) $ \\ 
$\left( 
\begin{array}{ccc}
0 & 0 & 0 \\ 
0 & \nu _{4} & 0 \\ 
0 & 0 & 0%
\end{array}
\right) $%
\end{tabular}
\\ \hline
$X_{\rho ^{ij}}$ & 
\begin{tabular}{c}
$-\frac{1}{6}\qquad \qquad \frac{1}{3}$ \\ 
$-\frac{2}{3}$%
\end{tabular}
& 
\begin{tabular}{c}
$-\frac{1}{3}\qquad \qquad \frac{1}{6}$ \\ 
$\frac{2}{3}$%
\end{tabular}
\\ \hline
\end{tabular}%
\end{equation*}

\begin{equation*}
\begin{tabular}{|c|c|c|}
\hline
& $\beta =\sqrt{3}$ & $\beta =-\sqrt{3}$ \\ \hline
$\left\langle \rho ^{ij}\right\rangle _{0}$ & 
\begin{tabular}{c}
$\left( 
\begin{array}{ccc}
0 & 0 & \nu _{3} \\ 
0 & \nu _{4} & 0 \\ 
\nu _{3} & 0 & 0%
\end{array}
\right) \left( 
\begin{array}{ccc}
\nu _{1} & 0 & 0 \\ 
0 & 0 & 0 \\ 
0 & 0 & 0%
\end{array}
\right) $ \\ 
$\left( 
\begin{array}{ccc}
0 & 0 & 0 \\ 
0 & 0 & \nu _{5} \\ 
0 & \nu _{5} & 0%
\end{array}
\right) $%
\end{tabular}
& 
\begin{tabular}{c}
$\left( 
\begin{array}{ccc}
\nu _{1} & 0 & 0 \\ 
0 & 0 & \nu _{5} \\ 
0 & \nu _{5} & 0%
\end{array}
\right) \left( 
\begin{array}{ccc}
0 & 0 & \nu _{3} \\ 
0 & 0 & 0 \\ 
\nu _{3} & 0 & 0%
\end{array}
\right) $ \\ 
$\left( 
\begin{array}{ccc}
0 & 0 & 0 \\ 
0 & \nu _{4} & 0 \\ 
0 & 0 & 0%
\end{array}
\right) $%
\end{tabular}
\\ \hline
$X_{\rho ^{ij}}$ & 
\begin{tabular}{c}
$0\qquad \qquad -1$ \\ 
$\frac{1}{2}$%
\end{tabular}
& 
\begin{tabular}{c}
$0\qquad \qquad -\frac{1}{2}$ \\ 
$1$%
\end{tabular}
\\ \hline
\end{tabular}%
\end{equation*}

\begin{table}[!tbh]
\begin{center}
$ 
\begin{tabular}{|c|c|}
\hline
& $\beta \neq \pm \frac{1}{\sqrt{3}};\pm \sqrt{3}$ \\ \hline
$\left\langle \rho ^{ij}\right\rangle _{0}$ & $\left( 
\begin{array}{ccc}
\nu _{1} & 0 & 0 \\ 
0 & 0 & 0 \\ 
0 & 0 & 0%
\end{array}
\right) \quad \left( 
\begin{array}{ccc}
0 & 0 & \nu _{3} \\ 
0 & 0 & 0 \\ 
\nu _{3} & 0 & 0%
\end{array}
\right) \quad \left( 
\begin{array}{ccc}
0 & 0 & 0 \\ 
0 & \nu _{4} & 0 \\ 
0 & 0 & 0%
\end{array}
\right) \quad \left( 
\begin{array}{ccc}
0 & 0 & 0 \\ 
0 & 0 & \nu _{5} \\ 
0 & \nu _{5} & 0%
\end{array}
\right) $ \\ \hline
$X_{\rho ^{ij}}$ & $-\frac{1}{2}-\frac{\beta }{2\sqrt{3}}\qquad \quad -\frac{%
1}{4}+\frac{\beta }{4\sqrt{3}}\quad \quad \qquad \frac{1}{2}-\frac{\beta }{2%
\sqrt{3}}\quad \quad \qquad \frac{1}{4}+\frac{\beta }{4\sqrt{3}}$ \\ \hline
\end{tabular}
$%
\end{center}
\caption{\textit{Vacuum alignments for the second SSB with Higgs sextets,
and for all values of $\protect\beta $.}}
\label{tab:tres}
\end{table}

In this way, we have found the Higgs bosons necessary to break the symmetry
according to the scheme $SU(3)_{L}\otimes U(1)_{X}\rightarrow
SU(2)_{L}\otimes U(1)_{Y}\rightarrow U(1)_{Q}$, and generate the masses of
the fermions and gauge fields. This choice fixes the values of the quantum
numbers $X$ and the value of $\beta $ needed to define the electric charge.
Among the possible solutions we have $\beta =-\sqrt{3}$, $\sqrt{3}$,$\ -1/%
\sqrt{3}$ and $1/\sqrt{3}$. They correspond to the models in Refs.\ \cite%
{ten}, \cite{eleven}, \cite{twelve}, and \cite{fourteen} respectively; the
Higgs sector of the latter will be developed in detailed in this paper. For $%
\beta =\pm 1/\sqrt{3}$ the models does not contain exotic electric charges.
For values of $\beta $ different from $\pm 1/\sqrt{3}$, exotic charges arise
i.e. electric charges different from $\pm 1$ and $0$.

\section{Higgs Potential\label{Higgs pot}}

\subsection{Three Higgs triplets}

The triplet field $\chi $ and the antitriplet $\chi ^{\ast }$ only introduce
VEV on the third component for the first transition, as it is indicated by
the solutions (\ref{triplewtecon1}) and (\ref{12}) respectively, it induces
the masses of the exotic fermionic components. In the second transition, it
is observed in the tables \ref{tab:uno} and \ref{tab:dos} that pairs of
solutions are obtained according to the value of $\beta $. A careful
analysis of such solutions shows that the pair of multiplets is necessary to
be able to give masses to the quarks of type up and down respectively%
\footnote{%
After the first breaking, the model 331 is reduced to the SM with two Higgs
doublets\cite{PLBus, Hunter}.}. Therefore, in the second transition is
necessary to introduce two triplets$\;\rho $ and $\eta $ associated to each
pair of solutions in the tables \ref{tab:uno} and \ref{tab:dos} \footnote{%
The choice of three triplets does not ensure necessarily the masses of all
leptons, an example is the model of Pisano and Pleitez, where an additional
sextet is needed \cite{ten}.}. In table \ref{tab:cuatro}, it is shown the
minimal contents of Higgs bosons necessary to break the symmetry for the
different values of $\beta $ and $X$.

\begin{table}[!tbh]
\begin{center}
$ 
\begin{tabular}{|l|}
\hline
1$^{st}$ SSB$\; 
\begin{tabular}{|c|c|c|c|}
\hline
& $\beta =\frac{1}{\sqrt{3}}$ & $\beta =-\frac{1}{\sqrt{3}}$ & $\beta \neq
\pm \frac{1}{\sqrt{3}}$ \\ \hline
$\left\langle \chi \right\rangle _{0}$ & $\left( 
\begin{array}{c}
0 \\ 
0 \\ 
\nu _{\chi _{3}}%
\end{array}
\right) $ & $\left( 
\begin{array}{c}
0 \\ 
0 \\ 
\nu _{\chi _{3}}%
\end{array}
\right) $ & $\;\left( 
\begin{array}{c}
0 \\ 
0 \\ 
\nu _{\chi _{3}}%
\end{array}
\right) \;$ \\ \hline
$X_{\chi }$ & $\frac{1}{3}$ & $-\frac{1}{3}$ & $\frac{\beta }{\sqrt{3}}$%
\end{tabular}
$ \\ \hline
2$^{nd}$ SSB$\; 
\begin{tabular}{|c|c|c|c|}
\hline
$\left\langle \rho \right\rangle _{0}$ & $\left( 
\begin{array}{c}
0 \\ 
\nu _{\rho _{2}} \\ 
\nu _{\rho _{3}}%
\end{array}
\right) $ & $\;\left( 
\begin{array}{c}
0 \\ 
\nu _{\rho _{2}} \\ 
0%
\end{array}
\right) \,$ & $\left( 
\begin{array}{c}
0 \\ 
\nu _{\rho _{2}} \\ 
0%
\end{array}
\right) $ \\ \hline
$X_{\rho }$ & $\frac{1}{3}$ & $\frac{2}{3}$ & $\frac{1}{2}-\frac{\beta }{2%
\sqrt{3}}$ \\ \hline
$\left\langle \eta \right\rangle _{0}$ & $\left( 
\begin{array}{c}
\nu _{\eta _{1}} \\ 
0 \\ 
0%
\end{array}
\right) $ & $\left( 
\begin{array}{c}
\nu _{\eta _{1}} \\ 
0 \\ 
\nu _{\eta _{3}}%
\end{array}
\right) $ & $\left( 
\begin{array}{c}
\nu _{\eta _{1}} \\ 
0 \\ 
0%
\end{array}
\right) $ \\ \hline
$X_{\eta }$ & $-\frac{2}{3}$ & $-\frac{1}{3}$ & $-\frac{1}{2}-\frac{\beta }{2%
\sqrt{3}}$ \\ \hline
\end{tabular}
$ \\ \hline
\end{tabular}
$%
\end{center}
\caption{\textit{Higgs triplets neccesary for the SSB scheme: $31\rightarrow
21\rightarrow 1$ for a fix value of $\protect\beta $.}}
\label{tab:cuatro}
\end{table}

In order to study the most general form of the Higgs potential, we take all
possible linear combinations among the three triplets forming quadratic,
cubic, and quartic products invariant under $SU(3)_{L}\otimes U(1)_{X}$. The
different potentials, renormalizable and $SU(3)_{L}\otimes U(1)_{X}$
invariant are

\begin{enumerate}
\item For $\beta =\frac{1}{\sqrt{3}}$%
\begin{align}
V_{higgs}&=\mu _{1}^{2}\chi ^{i}\chi _{i}+\mu _{2}^{2}\rho ^{i}\rho _{i}+\mu
_{3}^{2}\eta ^{i}\eta _{i}+\mu _{4}^{2}\left( \chi ^{i}\rho _{i}+h.c\right)
+f\left( \chi _{i}\rho _{j}\eta _{k}\varepsilon ^{ijk}+h.c\right)  \notag \\
&+\lambda _{1}(\chi ^{i}\chi _{i})^{2}+\lambda _{2}(\rho ^{i}\rho
_{i})^{2}+\lambda _{3}(\eta ^{i}\eta _{i})^{2}+\lambda _{4}\chi ^{i}\chi
_{i}\rho ^{j}\rho _{j}+\lambda _{5}\chi ^{i}\chi _{i}\eta ^{j}\eta _{j} 
\notag \\
&+\lambda _{6}\rho ^{i}\rho _{i}\eta ^{j}\eta _{j}+\lambda _{7}\chi ^{i}\eta
_{i}\eta ^{j}\chi _{j}+\lambda _{8}\chi ^{i}\rho _{i}\rho ^{j}\chi
_{j}+\lambda _{9}\eta ^{i}\rho _{i}\rho ^{j}\eta _{j}  \notag \\
&+\lambda _{10}\chi ^{i}\chi _{i}\left( \chi ^{j}\rho _{j}+h.c.\right)
+\lambda _{11}\rho ^{i}\rho _{i}\left( \rho ^{j}\chi _{j}+h.c.\right)
+\lambda _{12}\eta ^{i}\eta _{i}\left( \chi ^{j}\rho _{j}+h.c.\right)  \notag
\\
&+\lambda _{13}\left( \chi ^{i}\rho _{i}\chi ^{j}\rho _{j}+h.c.\right)
+\lambda _{14}\left( \eta ^{i}\chi _{i}\rho ^{j}\eta _{j}+h.c.\right) .
\label{21a}
\end{align}

\item For $\beta =-\frac{1}{\sqrt{3}}$%
\begin{align}
V_{higgs}& =\mu _{1}^{2}\chi ^{i}\chi _{i}+\mu _{2}^{2}\rho ^{i}\rho
_{i}+\mu _{3}^{2}\eta ^{i}\eta _{i}+\mu _{4}^{2}\left( \chi ^{i}\eta
_{i}+h.c\right) +f\left( \chi _{i}\rho _{j}\eta _{k}\varepsilon
^{ijk}+h.c\right)  \notag \\
& +\lambda _{1}(\chi ^{i}\chi _{i})^{2}+\lambda _{2}(\rho ^{i}\rho
_{i})^{2}+\lambda _{3}(\eta ^{i}\eta _{i})^{2}+\lambda _{4}\chi ^{i}\chi
_{i}\rho ^{j}\rho _{j}+\lambda _{5}\chi ^{i}\chi _{i}\eta ^{j}\eta _{j} 
\notag \\
& +\lambda _{6}\rho ^{i}\rho _{i}\eta ^{j}\eta _{j}+\lambda _{7}\chi
^{i}\eta _{i}\eta ^{j}\chi _{j}+\lambda _{8}\chi ^{i}\rho _{i}\rho ^{j}\chi
_{j}+\lambda _{9}\eta ^{i}\rho _{i}\rho ^{j}\eta _{j}  \notag \\
& +\lambda _{10}\chi ^{i}\chi _{i}\left( \chi ^{j}\eta _{j}+h.c.\right)
+\lambda _{11}\eta ^{i}\eta _{i}\left( \eta ^{j}\chi _{j}+h.c.\right)
+\lambda _{12}\rho ^{i}\rho _{i}\left( \chi ^{j}\eta _{j}+h.c.\right)  \notag
\\
& +\lambda _{13}\left( \chi ^{i}\eta _{i}\chi ^{j}\eta _{j}+h.c.\right)
+\lambda _{14}\left( \rho ^{i}\chi _{i}\eta ^{j}\rho _{j}+h.c.\right) .
\label{potbeta2}
\end{align}

\item For $\beta =\sqrt{3}$%
\begin{align}
V_{higgs}& =\mu _{1}^{2}\chi ^{i}\chi _{i}+\mu _{2}^{2}\rho ^{i}\rho
_{i}+\mu _{3}^{2}\eta ^{i}\eta _{i}+f\left( \chi _{i}\rho _{j}\eta
_{k}\varepsilon ^{ijk}+h.c\right) +\lambda _{1}(\chi ^{i}\chi
_{i})^{2}+\lambda _{2}(\rho ^{i}\rho _{i})^{2}  \notag \\
& +\lambda _{3}(\eta ^{i}\eta _{i})^{2}+\lambda _{4}\chi ^{i}\chi _{i}\rho
^{j}\rho _{j}+\lambda _{5}\chi ^{i}\chi _{i}\eta ^{j}\eta _{j}+\lambda
_{6}\rho ^{i}\rho _{i}\eta ^{j}\eta _{j}+\lambda _{7}\chi ^{i}\eta _{i}\eta
^{j}\chi _{j}  \notag \\
& +\lambda _{8}\chi ^{i}\rho _{i}\rho ^{j}\chi _{j}+\lambda _{9}\eta
^{i}\rho _{i}\rho ^{j}\eta _{j}+\lambda _{10}\left( \rho ^{i}\chi _{i}\rho
^{j}\eta _{j}+h.c.\right) .  \label{potbeta3}
\end{align}

\item For $\beta =-\sqrt{3}$%
\begin{align}
V_{higgs}& =\mu _{1}^{2}\chi ^{i}\chi _{i}+\mu _{2}^{2}\rho ^{i}\rho
_{i}+\mu _{3}^{2}\eta ^{i}\eta _{i}+f\left( \chi _{i}\rho _{j}\eta
_{k}\varepsilon ^{ijk}+h.c\right) +\lambda _{1}(\chi ^{i}\chi
_{i})^{2}+\lambda _{2}(\rho ^{i}\rho _{i})^{2}  \notag \\
& +\lambda _{3}(\eta ^{i}\eta _{i})^{2}+\lambda _{4}\chi ^{i}\chi _{i}\rho
^{j}\rho _{j}+\lambda _{5}\chi ^{i}\chi _{i}\eta ^{j}\eta _{j}+\lambda
_{6}\rho ^{i}\rho _{i}\eta ^{j}\eta _{j}+\lambda _{7}\chi ^{i}\eta _{i}\eta
^{j}\chi _{j}  \notag \\
& +\lambda _{8}\chi ^{i}\rho _{i}\rho ^{j}\chi _{j}+\lambda _{9}\eta
^{i}\rho _{i}\rho ^{j}\eta _{j}+\lambda _{10}\left( \eta ^{i}\chi _{i}\eta
^{j}\rho _{j}+h.c.\right) .  \label{potbeta4}
\end{align}

\item For $\beta $ arbitrary (different from $\pm \sqrt{3}$,\ $\pm 1/\sqrt{3}
$)
\end{enumerate}

\begin{eqnarray}
V_{higgs} &=&\mu _{1}^{2}\chi ^{i}\chi _{i}+\mu _{2}^{2}\rho ^{i}\rho
_{i}+\mu _{3}^{2}\eta ^{i}\eta _{i}+f\left( \chi _{i}\rho _{j}\eta
_{k}\varepsilon ^{ijk}+h.c.\right) +\lambda _{1}(\chi ^{i}\chi
_{i})^{2}+\lambda _{2}(\rho ^{i}\rho _{i})^{2}  \notag \\
&&+\lambda _{3}(\eta ^{i}\eta _{i})^{2}+\lambda _{4}\chi ^{i}\chi _{i}\rho
^{j}\rho _{j}+\lambda _{5}\chi ^{i}\chi _{i}\eta ^{j}\eta _{j}+\lambda
_{6}\rho ^{i}\rho _{i}\eta ^{j}\eta _{j}+\lambda _{7}\chi ^{i}\eta _{i}\eta
^{j}\chi _{j}  \notag \\
&&+\lambda _{8}\chi ^{i}\rho _{i}\rho ^{j}\chi _{j}+\lambda _{9}\eta
^{i}\rho _{i}\rho ^{j}\eta _{j}.  \label{gen b pot}
\end{eqnarray}

\subsection{Higgs sextet}

In some circumstances, the choice of three triplets may be not enough to
provide all leptons with masses \cite{ten, sixteen, seventeen}. Hence, an
additional sextet is considered, which should be compatible with any of the
solutions of the table \ref{tab:tres}. The choice of one of these solutions
depend on the fermionic sector to which we want to give masses. Once again,
when we evaluate all possible linear combinations among the four fields $%
\chi ,$ $\rho ,$ $\eta $ and the sextet $S$, additional terms are obtained
which should be added to the Higgs potentials found for the case of three
triplets, and according to the chosen value of $\beta $. They are shown in
tables \ref{tab:cinco}-\ref{tab:nueve}.

\begin{table}[!tbh]
\begin{center}
\begin{tabular}{|c|c|}
\hline
$X_{S}=\frac{1}{3}$ & 
\begin{tabular}{c}
$V(S)=\mu _{5}^{2}S^{ij}S_{ij}+f_{2}\left( \chi ^{i}S_{ij}\chi
^{j}+h.c\right) +\lambda _{15}\left( \chi _{i}S^{ij}\eta ^{k}\rho
^{l}\varepsilon _{jkl}+h.c\right) $ \\ 
$+S^{ij}S_{ij}\left( \lambda _{16}\chi ^{k}\chi _{k}+\lambda _{17}\rho
^{k}\rho _{k}+\lambda _{18}\eta ^{k}\eta _{k}\right) +\lambda
_{19}S^{ij}S_{ij}\left( \rho ^{k}\chi _{k}+h.c\right) $ \\ 
$+\lambda _{20}\chi ^{i}S_{ij}S^{jk}\chi _{k}+\lambda _{21}\rho
^{i}S_{ij}S^{jk}\rho _{k}+\lambda _{22}\eta ^{i}S_{ij}S^{jk}\eta
_{k}+\lambda _{23}\left( \rho ^{i}S_{ij}S^{jk}\chi _{k}+h.c\right) $ \\ 
$+\lambda _{24}(S^{ij}S_{ij})^{2}+\lambda _{25}S^{ij}S_{jk}S^{kl}S_{li}$%
\end{tabular}
\\ \hline
$X_{S}=-\frac{1}{6}$ & 
\begin{tabular}{c}
$V(S)=\mu _{5}^{2}S^{ij}S_{ij}+f_{2}\left( \eta ^{i}S_{ij}\rho
^{j}+h.c\right) +(\lambda _{15}\rho _{i}S^{ij}\rho ^{k}\chi ^{l}\varepsilon
_{jkl}$ \\ 
$+\lambda _{16}\eta _{i}S^{ij}\eta ^{k}\chi ^{l}\varepsilon
_{jkl}+h.c)+S^{ij}S_{ij}\left( \lambda _{17}\chi ^{k}\chi _{k}+\lambda
_{18}\rho ^{k}\rho _{k}+\lambda _{19}\eta ^{k}\eta _{k}\right) $ \\ 
$+\lambda _{20}S^{ij}S_{ij}\left( \rho ^{k}\chi _{k}+h.c\right) +\lambda
_{21}\chi ^{i}S_{ij}S^{jk}\chi _{k}+\lambda _{22}\rho ^{i}S_{ij}S^{jk}\rho
_{k}$ \\ 
$+\lambda _{23}\eta ^{i}S_{ij}S^{jk}\eta _{k}+\lambda _{24}\left( \rho
^{i}S_{ij}S^{jk}\chi _{k}+h.c\right) +\lambda
_{25}(S^{ij}S_{ij})^{2}+\lambda _{26}S^{ij}S_{jk}S^{kl}S_{li}$%
\end{tabular}
\\ \hline
$X_{S}=-\frac{2}{3}$ & 
\begin{tabular}{c}
$V(S)=\mu _{5}^{2}S^{ij}S_{ij}+f_{2}\left( \eta ^{i}S_{ij}\eta
^{j}+h.c\right) +\lambda _{15}\left( \eta _{i}S^{ij}\rho ^{k}\chi
^{l}\varepsilon _{jkl}+h.c\right) $ \\ 
+$S^{ij}S_{ij}\left( \lambda _{16}\chi ^{k}\chi _{k}+\lambda _{17}\rho
^{k}\rho _{k}+\lambda _{18}\eta ^{k}\eta _{k}\right) $ \\ 
$+\lambda _{19}\chi ^{i}S_{ij}S^{jk}\chi _{k}+\lambda _{20}\rho
^{i}S_{ij}S^{jk}\rho _{k}+\lambda _{21}\eta ^{i}S_{ij}S^{jk}\eta _{k}$ \\ 
$+\lambda _{22}(S^{ij}S_{ij})^{2}+\lambda _{23}S^{ij}S_{jk}S^{kl}S_{li}$%
\end{tabular}
\\ \hline
\end{tabular}%
\end{center}
\caption{\textit{Additional terms in the Higgs potential of Eq. ( \protect
\ref{21a}) for $\protect\beta =\frac{1}{\protect\sqrt{3}}$, when a Higgs
sextet is included.}}
\label{tab:cinco}
\end{table}
\begin{table}[!tbh]
\begin{center}
\begin{tabular}{|c|r|}
\hline
$X_{S}=-\frac{1}{3}$ & 
\begin{tabular}{c}
$V(S)=\mu _{5}^{2}S^{ij}S_{ij}+f_{2}\left( \chi ^{i}S_{ij}\chi
^{j}+h.c\right) +\lambda _{15}\left( \chi _{i}S^{ij}\eta ^{k}\rho
^{l}\varepsilon _{jkl}+h.c\right) $ \\ 
+$S^{ij}S_{ij}\left( \lambda _{16}\chi ^{k}\chi _{k}+\lambda _{17}\rho
^{k}\rho _{k}+\lambda _{18}\eta ^{k}\eta _{k}\right) $ \\ 
$\qquad \qquad +\lambda _{19}S^{ij}S_{ij}\left( \eta ^{k}\chi
_{k}+h.c\right) +\lambda _{20}\chi ^{i}S_{ij}S^{jk}\chi _{k}+\lambda
_{21}\rho ^{i}S_{ij}S^{jk}\rho _{k}$ \\ 
$+\lambda _{22}\eta ^{i}S_{ij}S^{jk}\eta _{k}+\lambda _{23}\left( \eta
^{i}S_{ij}S^{jk}\chi _{k}+h.c\right) +\lambda
_{24}(S^{ij}S_{ij})^{2}+\lambda _{25}S^{ij}S_{jk}S^{kl}S_{li}$%
\end{tabular}
\\ \hline
$X_{S}=\frac{1}{6}$ & \multicolumn{1}{|c|}{%
\begin{tabular}{c}
$V(S)=\mu _{5}^{2}S^{ij}S_{ij}+f_{2}\left( \eta ^{i}S_{ij}\rho
^{j}+h.c\right) +(\lambda _{15}\rho _{i}S^{ij}\rho ^{k}\chi ^{l}\varepsilon
_{jkl}$ \\ 
$+\lambda _{16}\eta _{i}S^{ij}\eta ^{k}\chi ^{l}\varepsilon
_{jkl}+h.c)+S^{ij}S_{ij}\left( \lambda _{17}\chi ^{k}\chi _{k}+\lambda
_{18}\rho ^{k}\rho _{k}+\lambda _{19}\eta ^{k}\eta _{k}\right) $ \\ 
$+\lambda _{20}S^{ij}S_{ij}\left( \eta ^{k}\chi _{k}+h.c\right) +\lambda
_{21}\chi ^{i}S_{ij}S^{jk}\chi _{k}+\lambda _{22}\rho ^{i}S_{ij}S^{jk}\rho
_{k}+\lambda _{23}\eta ^{i}S_{ij}S^{jk}\eta _{k}$ \\ 
$+\lambda _{24}\left( \eta ^{i}S_{ij}S^{jk}\chi _{k}+h.c\right) +\lambda
_{25}(S^{ij}S_{ij})^{2}+\lambda _{26}S^{ij}S_{jk}S^{kl}S_{li}$%
\end{tabular}%
} \\ \hline
$X_{S}=\frac{2}{3}$ & \multicolumn{1}{|c|}{%
\begin{tabular}{c}
$V(S)=\mu _{5}^{2}S^{ij}S_{ij}+f_{2}\left( \rho ^{i}S_{ij}\rho
^{j}+h.c\right) +\lambda _{15}\left( \rho _{i}S^{ij}\eta ^{k}\chi
^{l}\varepsilon _{jkl}+h.c\right) $ \\ 
+$S^{ij}S_{ij}\left( \lambda _{16}\chi ^{k}\chi _{k}+\lambda _{17}\rho
^{k}\rho _{k}+\lambda _{18}\eta ^{k}\eta _{k}\right) $ \\ 
$+\lambda _{19}\chi ^{i}S_{ij}S^{jk}\chi _{k}+\lambda _{20}\rho
^{i}S_{ij}S^{jk}\rho _{k}+\lambda _{21}\eta ^{i}S_{ij}S^{jk}\eta _{k}$ \\ 
$+\lambda _{22}(S^{ij}S_{ij})^{2}+\lambda _{23}S^{ij}S_{jk}S^{kl}S_{li}$%
\end{tabular}%
} \\ \hline
\end{tabular}%
\end{center}
\caption{\textit{Additional terms in the Higgs potential of Eq. (  \protect
\ref{potbeta2}) for $\protect\beta =-\frac{1}{\protect\sqrt{3}}$, when a
Higgs sextet is included.}}
\label{tab:seis}
\end{table}

\begin{table}[!tbh]
\begin{center}
\begin{tabular}{|c|c|}
\hline
$X_{S}=0$ & 
\begin{tabular}{c}
$V(S)=\mu _{5}^{2}S^{ij}S_{ij}+S^{ij}S_{ij}\left( \lambda _{15}\chi ^{k}\chi
_{k}+\lambda _{16}\rho ^{k}\rho _{k}+\lambda _{17}\eta ^{k}\eta _{k}\right) $
\\ 
$+\lambda _{18}\chi ^{i}S_{ij}S^{jk}\chi _{k}+\lambda _{19}\rho
^{i}S_{ij}S^{jk}\rho _{k}+\lambda _{20}\eta ^{i}S_{ij}S^{jk}\eta _{k}$ \\ 
$+\lambda _{21}(S^{ij}S_{ij})^{2}+\lambda _{22}S^{ij}S_{jk}S^{kl}S_{li}$%
\end{tabular}
\\ \hline
$X_{S}=-1$ & 
\begin{tabular}{c}
$V(S)=\mu _{5}^{2}S^{ij}S_{ij}+f_{2}\left( \eta ^{i}S_{ij}\eta
^{j}+h.c\right) +\lambda _{15}\left( \eta _{i}S^{ij}\rho ^{k}\chi
^{l}\varepsilon _{jkl}+h.c\right) $ \\ 
+$S^{ij}S_{ij}\left( \lambda _{17}\chi ^{k}\chi _{k}+\lambda _{18}\rho
^{k}\rho _{k}+\lambda _{19}\eta ^{k}\eta _{k}\right) +\lambda _{20}\chi
^{i}S_{ij}S^{jk}\chi _{k}$ \\ 
$+\lambda _{21}\rho ^{i}S_{ij}S^{jk}\rho _{k}+\lambda _{22}\eta
^{i}S_{ij}S^{jk}\eta _{k}+\lambda _{23}(S^{ij}S_{ij})^{2}+\lambda
_{24}S^{ij}S_{jk}S^{kl}S_{li}$%
\end{tabular}
\\ \hline
$X_{S}=\frac{1}{2}$ & 
\begin{tabular}{c}
$V(S)=\mu _{5}^{2}S^{ij}S_{ij}+f_{2}\left( \rho ^{i}S_{ij}\chi
^{j}+h.c\right) +(\lambda _{15}\chi _{i}S^{ij}\eta ^{k}\chi ^{l}\varepsilon
_{jkl}$ \\ 
$+\lambda _{16}\rho _{i}S^{ij}\eta ^{k}\rho ^{l}\varepsilon
_{jkl}+h.c)+S^{ij}S_{ij}\left( \lambda _{17}\chi ^{k}\chi _{k}+\lambda
_{18}\rho ^{k}\rho _{k}+\lambda _{19}\eta ^{k}\eta _{k}\right) $ \\ 
$+\lambda _{20}\chi ^{i}S_{ij}S^{jk}\chi _{k}+\lambda _{21}\rho
^{i}S_{ij}S^{jk}\rho _{k}+\lambda _{22}\eta ^{i}S_{ij}S^{jk}\eta _{k}$ \\ 
$+\lambda _{23}(S^{ij}S_{ij})^{2}+\lambda _{24}S^{ij}S_{jk}S^{kl}S_{li}$%
\end{tabular}
\\ \hline
\end{tabular}%
\end{center}
\caption{\textit{Additional terms in the Higgs potential of Eq. ( \protect
\ref{potbeta3}) for $\protect\beta =\protect\sqrt{3}$, when a Higgs sextet
is included.}}
\label{tab:siete}
\end{table}

\begin{table}[!tbh]
\begin{center}
\begin{tabular}{|c|c|}
\hline
$X_{S}=0$ & 
\begin{tabular}{c}
$V(S)=\mu _{5}^{2}S^{ij}S_{ij}+S^{ij}S_{ij}\left( \lambda _{15}\chi ^{k}\chi
_{k}+\lambda _{16}\rho ^{k}\rho _{k}+\lambda _{17}\eta ^{k}\eta _{k}\right) $
\\ 
$+\lambda _{18}\chi ^{i}S_{ij}S^{jk}\chi _{k}+\lambda _{19}\rho
^{i}S_{ij}S^{jk}\rho _{k}+\lambda _{20}\eta ^{i}S_{ij}S^{jk}\eta _{k}$ \\ 
$+\lambda _{21}(S^{ij}S_{ij})^{2}+\lambda _{22}S^{ij}S_{jk}S^{kl}S_{li}$%
\end{tabular}
\\ \hline
$X_{S}=1$ & 
\begin{tabular}{c}
$V(S)=\mu _{5}^{2}S^{ij}S_{ij}+f_{2}\left( \rho ^{i}S_{ij}\rho
^{j}+h.c\right) +\lambda _{15}\left( \rho _{i}S^{ij}\eta ^{k}\chi
^{l}\varepsilon _{jkl}+h.c\right) $ \\ 
+$S^{ij}S_{ij}\left( \lambda _{17}\chi ^{k}\chi _{k}+\lambda _{18}\rho
^{k}\rho _{k}+\lambda _{19}\eta ^{k}\eta _{k}\right) $ \\ 
$+\lambda _{20}\chi ^{i}S_{ij}S^{jk}\chi _{k}+\lambda _{21}\rho
^{i}S_{ij}S^{jk}\rho _{k}+\lambda _{22}\eta ^{i}S_{ij}S^{jk}\eta _{k}$ \\ 
$+\lambda _{23}(S^{ij}S_{ij})^{2}+\lambda _{24}S^{ij}S_{jk}S^{kl}S_{li}$%
\end{tabular}
\\ \hline
$X_{S}=-\frac{1}{2}$ & 
\begin{tabular}{c}
$V(S)=\mu _{5}^{2}S^{ij}S_{ij}+f_{2}\left( \eta ^{i}S_{ij}\chi
^{j}+h.c\right) +(\lambda _{15}\chi _{i}S^{ij}\rho ^{k}\chi ^{l}\varepsilon
_{jkl}$ \\ 
$+\lambda _{16}\eta _{i}S^{ij}\eta ^{k}\rho ^{l}\varepsilon
_{jkl}+h.c)+S^{ij}S_{ij}\left( \lambda _{17}\chi ^{k}\chi _{k}+\lambda
_{18}\rho ^{k}\rho _{k}+\lambda _{19}\eta ^{k}\eta _{k}\right) $ \\ 
$+\lambda _{20}\chi ^{i}S_{ij}S^{jk}\chi _{k}+\lambda _{21}\rho
^{i}S_{ij}S^{jk}\rho _{k}+\lambda _{22}\eta ^{i}S_{ij}S^{jk}\eta _{k}$ \\ 
$+\lambda _{23}(S^{ij}S_{ij})^{2}+\lambda _{24}S^{ij}S_{jk}S^{kl}S_{li}$%
\end{tabular}
\\ \hline
\end{tabular}%
\end{center}
\caption{\textit{Additional terms in the Higgs potential of Eq. (  \protect
\ref{potbeta4}) for $\protect\beta =-\protect\sqrt{3}$, when a Higgs sextet
is included.}}
\label{tab:ocho}
\end{table}
\begin{table}[!tbh]
\begin{center}
$ 
\begin{tabular}{|c|r|}
\hline
$\beta $ arbitrary & 
\begin{tabular}{c}
$V(S)=\mu _{5}^{2}S^{ij}S_{ij}+S^{ij}S_{ij}\left( \lambda _{15}\chi ^{k}\chi
_{k}+\lambda _{16}\rho ^{k}\rho _{k}+\lambda _{17}\eta ^{k}\eta _{k}\right) $
\\ 
$+\lambda _{18}\chi ^{i}S_{ij}S^{jk}\chi _{k}+\lambda _{19}\rho
^{i}S_{ij}S^{jk}\rho _{k}+\lambda _{20}\eta ^{i}S_{ij}S^{jk}\eta _{k}$ \\ 
$+\lambda _{21}(S^{ij}S_{ij})^{2}+\lambda _{22}S^{ij}S_{jk}S^{kl}S_{li}$%
\end{tabular}
\\ \hline
\end{tabular}
$%
\end{center}
\caption{\textit{Additional terms in the Higgs potential of Eq. ( \protect
\ref{gen b pot}) for $\protect\beta \neq \pm \frac{1}{\protect\sqrt{3}}$ and 
$\protect\beta \neq \pm \protect\sqrt{3}$, when a Higgs sextet is included.}}
\label{tab:nueve}
\end{table}

\section{331 model with $\protect\beta =1/\protect\sqrt{3}$}

\subsection{Fermionic content}

Before describing in detail the 331 model with $\beta =1/\sqrt{3}$, let us
check that it could provide a realistic scenario. As we saw in section \ref%
{S repres}, we found the Higgs structures and vacuum alignments that satisfy
the gauge invariance, the symmetry breaking pattern and the adequate
assignment of masses. The quantum numbers associated to $U\left( 1\right)
_{X}$ for three Higgs triplets in the specific case of $\beta =1/\sqrt{3}$,
can be determined from Eq. (\ref{triplewtecon1}) for the triplet associated
to the first transition and from table \ref{tab:uno} for the two triplets of
the second transition. On the other hand, gauge invariance in the Yukawa
sector provides some relations among the $X_{\Phi }$ and $X_{f}$ numbers
associated to $U\left( 1\right) _{X}$, corresponding to scalars and fermions
respectively. Further, cancelation of fermionic anomalies and the
requirement that the first two components of the fermion triplets of $%
SU\left( 3\right) _{L}\ $transform as doublets of the SM, provide the
complete requirements to write down the quantum number assignments for the
fermionic sector of the model. From the fermionic content described in table %
\ref{tab:fermionnum} with $\beta =1/\sqrt{3}$ we get the quantum numbers of
table \ref{tab:fermionnum2} for a one family framework. This table shows us
that the standard model fermions acquire the appropiate quantum numbers and
charges, and that the exotic fermions do not have exotic charges either. 
\begin{table}[!tbh]
\begin{center}
$%
\begin{tabular}{||c||c||c||c||}
\hline\hline
$representation$ & $Q_{\psi }$ & $Y_{\psi }$ & $X_{\psi }$ \\ \hline\hline
$%
\begin{tabular}{c}
$q_{L}=\left( 
\begin{array}{c}
u \\ 
d \\ 
J%
\end{array}%
\right) _{L}:3$ \\ 
\\ 
$q_{R}^{(1)}=u_{R}:1$ \\ 
$q_{R}^{(2)}=d_{R}:1$ \\ 
$q_{R}^{(3)}=J_{R}:1$%
\end{tabular}%
\ $ & 
\begin{tabular}{c}
$\left( 
\begin{array}{c}
\frac{2}{3} \\ 
-\frac{1}{3} \\ 
-\frac{1}{3}%
\end{array}%
\right) $ \\ 
\\ 
$\frac{2}{3}$ \\ 
$-\frac{1}{3}$ \\ 
$-\frac{1}{3}$%
\end{tabular}
& 
\begin{tabular}{c}
$\left( 
\begin{array}{c}
\frac{1}{6} \\ 
\frac{1}{6} \\ 
-\frac{1}{3}%
\end{array}%
\right) $ \\ 
\\ 
$\frac{2}{3}$ \\ 
$-\frac{1}{3}$ \\ 
$-\frac{1}{3}$%
\end{tabular}
& 
\begin{tabular}{c}
\\ 
$X_{q}^{L}=0$ \\ 
\\ 
\\ 
$X_{u}^{R}=\frac{2}{3}$ \\ 
$X_{d}^{R}=-\frac{1}{3}$ \\ 
$X_{J}^{R}=-\frac{1}{3}$%
\end{tabular}
\\ \hline\hline
$\ell _{L}^{(1)}=\left( 
\begin{array}{c}
e^{-} \\ 
-\nu _{e} \\ 
E_{1}^{0}%
\end{array}%
\right) _{L}:3^{\ast }$ & $\left( 
\begin{array}{c}
-1 \\ 
0 \\ 
0%
\end{array}%
\right) $ & $\left( 
\begin{array}{c}
-\frac{1}{2} \\ 
-\frac{1}{2} \\ 
0%
\end{array}%
\right) $ & $X_{\ell ^{(1)}}^{L}=\frac{1}{3}$ \\ \hline\hline
$\ell _{L}^{(2)}=\left( 
\begin{array}{c}
E_{2}^{0} \\ 
-E_{1}^{+} \\ 
e^{+}%
\end{array}%
\right) _{L}:3^{\ast }$ & $\left( 
\begin{array}{c}
0 \\ 
1 \\ 
1%
\end{array}%
\right) $ & $\left( 
\begin{array}{c}
\frac{1}{2} \\ 
\frac{1}{2} \\ 
1%
\end{array}%
\right) $ & $X_{\ell ^{(2)}}^{L}=-\frac{2}{3}$ \\ \hline\hline
$\ell _{L}^{(3)}=\left( 
\begin{array}{c}
E_{1}^{-} \\ 
-E_{3}^{0} \\ 
E_{4}^{0}%
\end{array}%
\right) _{L}:3^{\ast }$ & $\left( 
\begin{array}{c}
-1 \\ 
0 \\ 
0%
\end{array}%
\right) $ & $\left( 
\begin{array}{c}
-\frac{1}{2} \\ 
-\frac{1}{2} \\ 
0%
\end{array}%
\right) $ & $X_{\ell ^{(3)}}^{L}=\frac{1}{3}$ \\ \hline\hline
\end{tabular}%
$%
\end{center}
\caption{\textit{Particle content for the fermionic sector of the 331 model
with $\protect\beta =1/\protect\sqrt{3}$ for a one family scenario.}}
\label{tab:fermionnum2}
\end{table}

\subsection{Scalar mass spectrum for $\protect\beta =1/\protect\sqrt{3}$%
\label{S masses}}

The masses of the scalar sector arise from the implementation of the SSB
over the Higgs potentials. A detailed calculation of the scalar spectrum is
done in Ref. \cite{long} in models with $\beta =-\sqrt{3}$ and $\beta =-%
\frac{1}{\sqrt{3}},$ that coincide for these values of $\beta $, with the
solutions of table \ref{tab:cuatro} for the triplets, and of table \ref%
{tab:tres} for the sextet. In this article, we examine in detail a model
with $\beta =\frac{1}{\sqrt{3}},$ which has no exotic charges. With three
Higgs triplets, the most general potential is given by Eq. (\ref{21a}),
where the solution in table \ref{tab:cuatro} is presented more explicitly in
table \ref{tab:diez}.

\begin{table}[!tbh]
\begin{center}
$ 
\begin{tabular}{|c|c|c|c|c|}
\hline
& $Q_{\Phi }$ & $Y_{\Phi }$ & $X_{\Phi }$ & $\left\langle \Phi \right\rangle
_{0}$ \\ \hline
$\chi =\left( 
\begin{array}{c}
\chi ^{\pm } \\ 
\xi _{\chi ^{0}}\pm i\zeta _{\chi ^{0}} \\ 
\xi _{\chi }\pm i\zeta _{\chi }%
\end{array}
\right) $ & $\left( 
\begin{array}{c}
\pm 1 \\ 
0 \\ 
0%
\end{array}
\right) $ & $\left( 
\begin{array}{c}
\pm \frac{1}{2} \\ 
\pm \frac{1}{2} \\ 
0%
\end{array}
\right) $ & $\frac{1}{3}$ & $\left( 
\begin{array}{c}
0 \\ 
0 \\ 
\nu _{\chi ^{3}}%
\end{array}
\right) $ \\ \hline
$\rho =\left( 
\begin{array}{c}
\rho ^{\pm } \\ 
\xi _{\rho ^{2}}\pm i\zeta _{\rho ^{2}} \\ 
\xi _{\rho ^{3}}\pm i\zeta _{\rho ^{3}}%
\end{array}
\right) $ & $\left( 
\begin{array}{c}
\pm 1 \\ 
0 \\ 
0%
\end{array}
\right) $ & $\left( 
\begin{array}{c}
\pm \frac{1}{2} \\ 
\pm \frac{1}{2} \\ 
\,0%
\end{array}
\right) $ & $\frac{1}{3}$ & $\left( 
\begin{array}{c}
0 \\ 
\nu _{\rho ^{2}} \\ 
\nu _{\rho ^{3}}%
\end{array}
\right) $ \\ \hline
$\eta =\left( 
\begin{array}{c}
\xi _{\eta }\pm i\zeta _{\eta } \\ 
\eta ^{2\mp } \\ 
\chi ^{3\mp }%
\end{array}
\right) $ & $\left( 
\begin{array}{c}
0 \\ 
\mp 1 \\ 
\mp 1%
\end{array}
\right) $ & $\left( 
\begin{array}{c}
\mp \frac{1}{2} \\ 
\mp \frac{1}{2} \\ 
\,\mp 1%
\end{array}
\right) $ & $-\frac{2}{3}$ & $\left( 
\begin{array}{c}
\nu _{\eta } \\ 
0 \\ 
0%
\end{array}
\right) $ \\ \hline
\end{tabular}
$%
\end{center}
\caption{\textit{Quantum numbers of three scalar triplets for $\protect\beta %
=\frac{1}{\protect\sqrt{3}}$}}
\label{tab:diez}
\end{table}

\noindent The conditions for the minimum of the potential are given by 
\begin{equation}
\frac{\partial \left\langle V_{higgs}\right\rangle }{\partial \nu _{\chi }}%
=0\;,\quad \frac{\partial \left\langle V_{higgs}\right\rangle }{\partial \nu
_{\rho ^{2}}}=0\;,\quad \frac{\partial \left\langle V_{higgs}\right\rangle }{%
\partial \nu _{\rho ^{3}}}=0\;,\quad \frac{\partial \left\langle
V_{higgs}\right\rangle }{\partial \nu _{\eta }}=0,  \label{22}
\end{equation}%
and evaluated when all fields are equal to zero, we get 
\begin{eqnarray}
\mu _{1}^{2} &=&-2\lambda _{1}\nu _{\chi }^{2}-\lambda _{4}\nu _{\eta
}^{2}-\lambda _{5}\left( \nu _{\rho ^{2}}^{2}+\nu _{\rho ^{3}}^{2}\right)
-2\lambda _{10}\nu _{\chi }\nu _{\rho ^{3}}-f\left( \frac{\nu _{\eta }\nu
_{\rho ^{3}}^{2}}{\nu _{\chi }\nu _{\rho ^{2}}}+\frac{\nu _{\eta }\nu _{\rho
^{2}}}{\nu _{\chi }}\right) ,  \notag \\
\mu _{2}^{2} &=&-2\lambda _{2}\left( \nu _{\rho ^{2}}^{2}+\nu _{\rho
^{3}}^{2}\right) -\lambda _{5}\nu _{\chi }^{2}-\lambda _{6}\nu _{\eta
}^{2}-2\lambda _{11}\nu _{\chi }\nu _{\rho ^{3}}-f\frac{\nu _{\eta }\nu
_{\chi }}{\nu _{\rho ^{2}}},  \notag \\
\mu _{3}^{2} &=&-2\lambda _{3}\nu _{\eta }^{2}-\lambda _{4}\nu _{\chi
}^{2}-\lambda _{6}\left( \nu _{\rho ^{2}}^{2}+\nu _{\rho ^{3}}^{2}\right)
-2\lambda _{12}\nu _{\chi }\nu _{\rho ^{3}}-f\frac{\nu _{\rho ^{2}}\nu
_{\chi }}{\nu _{\eta }},  \notag \\
\mu _{4}^{2} &=&-\lambda _{8}\nu _{\chi }\nu _{\rho ^{3}}-\lambda _{10}\nu
_{\chi }^{2}-\lambda _{11}\left( \nu _{\rho ^{2}}^{2}+\nu _{\rho
^{3}}^{2}\right) -\lambda _{12}\nu _{\eta }^{2}-2\lambda _{13}\nu _{\chi
}\nu _{\rho ^{3}}+f\frac{\nu _{\eta }\nu _{\rho ^{3}}}{\nu _{\rho ^{2}}}.
\end{eqnarray}

These parameters are again replaced in the Higgs potential. If the potential
is derived twice and then evaluated for all scalar fields such that $%
\widehat{\Phi }_{i}=0$, we find the square mass terms. For the neutral
scalar fields we have that 
\begin{equation}
M_{\widehat{\Phi }_{i}\widehat{\Phi }_{j}}^{2}=\left. \frac{\partial
^{2}V_{higgs}}{\partial \widehat{\Phi }_{i}\partial \widehat{\Phi }_{j}}%
\right] _{\widehat{\Phi }_{i}=0}
\end{equation}%
with $\widehat{\Phi }_{i}=\zeta _{\chi ^{0}},\zeta _{\chi },\zeta _{\rho
^{2}},\zeta _{\rho ^{3}},\zeta _{\eta }$ (pseudoscalars) and $\widehat{\Phi }%
_{i}=\xi _{\chi ^{0}},\xi _{\chi },\xi _{\rho ^{2}},\xi _{\rho ^{3}},\xi
_{\eta }$ (real). For the charged components we have 
\begin{equation}
M_{\widehat{\Phi }_{i}^{\ast }\widehat{\Phi }_{j}}^{2}=\left. \frac{\partial
^{2}V_{higgs}}{\partial \widehat{\Phi }_{i}^{\ast }\partial \widehat{\Phi }%
_{j}}\right] _{\widehat{\Phi }_{i}=0}
\end{equation}%
with $\widehat{\Phi }_{i}=\chi ^{+},\rho ^{+},\eta ^{2+},\eta ^{3+}$. From
the equations above, we obtain the mass matrices $M_{\zeta \zeta }^{2}$ for
the imaginary sector, $M_{\xi \xi }^{2}$ for the scalar real sector, and $%
M_{\phi ^{\pm }}^{2}$ for the charged scalar sector. They are written
explicitly in appendix \ref{ap:S masses} Eqs. (\ref{imaginario}), (\ref{real}%
) and (\ref{cargado}), respectively.

\subsubsection{Diagonalization of the charged sector}

To obtain the physical spectrum of scalar particles, we should diagonalize
the mass matrices. For the matrix $M_{\phi ^{\pm }}^{2}$ in (\ref{cargado})
it is obtain that $\det $($M_{\phi ^{\pm }}^{2})=0,$ which guarantees null
eigenvalues associated to the would be Goldstone bosons. The eigenvalue
equation is 
\begin{equation}
\det (M_{\phi ^{\pm }}^{2}-P_{i}I)=0.  \label{24}
\end{equation}%
Resolving this equation, a degeneration with $P_{1}=P_{2}=0,$ is observed;
generating two would be Goldstone bosons associated to two charged gauge
fields that acquire mass $W_{\mu }^{+},K_{\mu }^{+}$. The eigenvectors $%
V_{i} $ associated to each eigenvalue are obtained from 
\begin{equation}
\left( M_{\phi ^{\pm }}^{2}-IP_{i}\right) \cdot V_{i}=0,  \label{25}
\end{equation}%
When solving this equation for the null eigenvalue modes (see appendix \ref%
{ap:diagcar}), and using the basis $\chi ^{\pm },\rho ^{\pm },\eta ^{2\pm
},\eta ^{3\pm }$, it leads to the following combinations, except for a
normalization factor 
\begin{eqnarray}
\phi _{1}^{\pm } &\approx &\left( -\nu _{\rho ^{2}}\rho ^{\pm }+\nu _{\eta
}\eta ^{2\pm }\right) ,  \notag \\
\phi _{2}^{\pm } &\approx &\left( \nu _{\chi }\chi ^{\pm }+\nu _{\rho
^{3}}\rho ^{\pm }-\nu _{\eta }\eta ^{3\pm }\right) ,  \label{26a}
\end{eqnarray}

In the $R_{\zeta }$ gauge, $\mid \partial _{\mu }X^{\mu }-\alpha
M_{X}G_{X}\mid ^{2}/\alpha $, the would be Goldstone boson $G_{X}$ of the
gauge field $X_{\mu }$ can be defined if we demand from the bilinear term of
the gauge fixing to be canceled with the bilinear term coming from the
kinetic scalar Lagrangian. In this case we observe that the equation above
coincides directly with the first two relations of the Eq. (\ref{B9}) in
appendix \ref{ap:acoplesGg}; where the would be Goldstone bosons of the
theory have been obtained from the bilinear term of scalar-gauge fields that
appear in the kinetic term of the scalar fields. Therefore, it is deduced
that $\phi _{1}^{\pm }$ endows the electroweak charged gauge bosons $W_{\mu
}^{\pm }$\ with mass; while $\phi _{2}^{\pm }$ gives masses to the exotic
gauge bosons $K_{\mu }^{\pm }.$

The rest of the eigenvalues are complicate expressions no easy to visualize
directly. To find a solution with a clear significance, we should bear in
mind that the exotic particles (non observed phenomenologically at low
energies) acquire their masses in the first transition of SSB, which happens
at very high energies respect to the second transition. It means that\ \cite%
{long}: 
\begin{equation}
\left\langle \chi \right\rangle _{0}\gg \left\langle \rho \right\rangle
_{0},\left\langle \eta \right\rangle _{0}.  \label{27}
\end{equation}%
Under this approximation, it is valid to keep only terms involving $\nu
_{\chi }$ in the matrix (\ref{cargado}). If the aproximation of only
quadratic terms $\nu _{\chi }^{2}$ is held, it is obtain the matrix shown in
appendix \ref{ap:diagcar} Eq. (\ref{47}). As it is observed, this
approximation cancels all the components, except the term in the position $%
M_{\phi ^{\pm }}^{2}(4,4)$, corresponding to the field $\eta ^{3\pm },$ from
which information about one of the non-zero eigenvalues is lost. To obtain a
better approximation we should allow additional terms but without demanding
linear orders in $\nu _{\chi }$. One way to have this, is by considering the
coefficient $f$ of the cubic term in the scalar potential eq. (\ref{21a})
under the following approximation \cite{long}: 
\begin{equation}
\left\vert f\right\vert \approx \nu _{\chi }.  \label{28}
\end{equation}%
It means that terms of the form $f\nu _{\chi }$ in the mass matrix Eq. (\ref%
{cargado}) are of the order $O\left( \nu _{\chi }^{2}\right) $, and should
be preserved in the aproximation, from which we get the matrix of Eq. (\ref%
{48}). As it was mentioned in Ref. \cite{long}, \ the approximation (\ref{28}%
) avoids the introduction of another mass scale apart from the ones defined
by the two transitions in Eq. (\ref{res}). Effectively, when we study the
first transition only, we obtain Higgs square masses of the order of $\nu
_{\chi }^{2}$ and $f\nu _{\chi }$, which can be considered of the same order
of magnitude if $f\sim \nu _{\chi }$.

In the matrix (\ref{48}), it is observed that the component $\eta ^{3\pm }$
decouples, and the mass matrix is reduced to $M_{\rho \eta ^{2}}^{2}$ Eq. (%
\ref{49}) of appendix \ref{ap:diagcar}. Getting the eigenvalue $P_{3}=$ $%
\lambda _{7}\nu _{\chi }^{2}-f\nu _{\chi }\frac{\nu _{\rho ^{2}}}{\nu _{\eta
}}$, i.e.,

\begin{equation}
h_{1}^{\pm }=\eta ^{3\pm },\;\;\;M_{h_{1}^{\pm }}^{2}\approx \lambda _{7}\nu
_{\chi }^{2}-f\nu _{\chi }\frac{\nu _{\rho ^{2}}}{\nu _{\eta }}.  \label{28b}
\end{equation}%
and the last eigenvalue is 
\begin{equation}
P_{4}\approx -f\nu _{\chi }\left( \frac{\nu _{\eta }}{\nu _{\rho ^{2}}}+%
\frac{\nu _{\rho ^{2}}}{\nu _{\eta }}\right) ,  \label{29}
\end{equation}%
whose eigenvector in the basis $\left( \rho ^{\pm },\eta ^{2\pm }\right) $ is%
$:$

\begin{equation}
h_{2}^{\pm }\approx C_{\theta }\rho ^{\pm }+S_{\theta }\eta ^{2\pm },
\label{31a}
\end{equation}
with 
\begin{equation}
S_{\theta }=\sin \theta =\frac{\nu _{\rho ^{2}}}{\sqrt{\nu _{\rho
^{2}}^{2}+\nu _{\eta }^{2}}},\quad C_{\theta }=\cos \theta =\frac{\nu _{\eta
}}{\sqrt{\nu _{\rho ^{2}}^{2}+\nu _{\eta }^{2}}}.  \label{32}
\end{equation}

Using the hierarchy of the VEV's in Eq. (\ref{27}) we finally find the
particles spectrum of the scalar charged sector and their masses which we
summarize in table \ref{tab:once}. The fields $\phi _{1}^{\pm }$ and $\phi
_{2}^{\pm }$ correspond to four charged and massless would-be Goldstone
bosons, associated to four charged gauge bosons $W_{\mu }^{\pm }$ and $%
K_{\mu }^{\pm }$ respectively; while $h_{1}^{\pm }$ and $h_{2}^{\pm }$
appears in the spectrum as four charged Higgs bosons with the masses
indicated in table \ref{tab:once}, of the order of the scale of the first
SSB. The Higgs bosons $h_{2}^{\pm }$ have real and positive masses if $f<0.$

\begin{table}[!tbh]
\begin{center}
$ 
\begin{tabular}{||c||c||c||}
\hline\hline
Charged scalars & Squared masses & Features \\ \hline\hline
$\phi _{1}^{\pm }=-S_{\theta }\rho ^{\pm }+C_{\theta }\eta ^{2\pm },$ & $%
M_{\phi _{1}^{\pm }}^{2}=0$ & 
\begin{tabular}{l}
Goldstone \\ 
associated to $W_{\mu }^{\pm }$%
\end{tabular}
\\ \hline\hline
$\phi _{2}^{\pm }\simeq \chi ^{\pm }$ & $M_{\phi _{2}^{\pm }}^{2}=0$ & $%
\begin{tabular}{l}
Goldstone \\ 
associated to $K_{\mu }^{\pm }$%
\end{tabular}
$ \\ \hline\hline
$h_{1}^{\pm }\simeq \eta ^{3\pm }$ & $M_{h_{1}^{\pm }}^{2}\simeq \lambda
_{7}\nu _{\chi }^{2}-f\nu _{\chi }\frac{\nu _{\rho ^{2}}}{\nu _{\eta }}$ & 
Higgs \\ \hline\hline
$h_{2}^{\pm }\simeq C_{\theta }\rho ^{\pm }+S_{\theta }\eta ^{2\pm }$ & $%
M_{h_{2}^{\pm }}^{2}\simeq -f\nu _{\chi }\left( \frac{\nu _{\eta }}{\nu
_{\rho ^{2}}}+\frac{\nu _{\rho ^{2}}}{\nu _{\eta }}\right) $ & Higgs \\ 
\hline\hline
\end{tabular}
$%
\end{center}
\caption{\textit{Physical spectrum of charged scalar particles, for three
Higgs triplets and $\protect\beta =1/\protect\sqrt{3}$.}}
\label{tab:once}
\end{table}

\subsubsection{Diagonalization of the imaginary sector}

Now we proceed to diagonalize the matrix $M_{\zeta \zeta }^{2}$ of Eq. (\ref%
{imaginario}) appendix \ref{ap:S masses}, which has null determinant
associated to the would be Goldstone bosons of the neutral massive gauge
fields. Solving the eigenvalues equation for $M_{\zeta \zeta }^{2}$ we get
the eigenvalues indicated in appendix \ref{ap:diagim} Eq. (\ref{50}); three
of them are degenerate with eigenvalue zero.

Since the triplets of the scalar fields $\chi $ and $\rho $ have the same
quantum numbers they can be rotated to eliminate one VEV \cite{georgi}. With
this basis it is simpler to find the mass eigenstates of the scalar sector 
\begin{eqnarray}
&&\left( 
\begin{tabular}{l}
$0$ \\ 
$\nu _{\chi ^{2}}^{\prime }$ \\ 
$\nu _{\chi ^{3}}^{\prime }$%
\end{tabular}%
\ \right) \rightarrow \left( 
\begin{tabular}{l}
$0$ \\ 
$\nu _{\chi ^{2}}$ \\ 
$\nu _{\chi ^{3}}$%
\end{tabular}%
\ \right) \ \ ,  \notag \\
&&\left( 
\begin{tabular}{l}
$0$ \\ 
$\nu _{\rho ^{2}}^{\prime }$ \\ 
$\nu _{\rho ^{3}}^{\prime }$%
\end{tabular}%
\ \right) \rightarrow \left( 
\begin{tabular}{l}
$0$ \\ 
$\nu _{\rho ^{2}}$ \\ 
$0$%
\end{tabular}%
\ \right) .  \label{rotVEV}
\end{eqnarray}

The solution of the characteristic equation generates the eigenvectors
written in their complete form in Eq. (\ref{51}). For the first three
degenerate eigenvalues $P_{1}=P_{2}=P_{3}=0,$ we have the corresponding
eigenvectors associated to the three would be Goldstone bosons. Using the
basis of the matrix $M_{\zeta \zeta }^{2}$ Eq. (\ref{imaginario}), we have 
\begin{eqnarray}
\phi _{1}^{0} &=&N_{1}^{0}\left( \nu _{\rho ^{2}}\zeta _{\rho ^{3}}+\nu
_{\chi }\zeta _{\chi ^{0}}\right) ,  \notag \\
\phi _{2}^{0} &=&N_{2}^{0}\left( \nu _{\chi }\zeta _{\chi }-\nu _{\eta
}\zeta _{\eta }\right) ,  \notag \\
\phi _{3}^{0} &=&N_{3}^{0}\left( -\nu _{\rho ^{2}}\zeta _{\rho ^{2}}+\nu
_{\eta }\zeta _{\eta }\right)  \label{32c}
\end{eqnarray}%
When we compare the first and the third of these combinations with the ones
obtained in the appendix \ref{ap:acoplesGg} Eq.(\ref{B10}), they could be
identified with $\phi _{1}^{0}$ and $\phi _{3}^{0}$ which are associated to
the longitudinal modes of the gauge fields $\overline{\widetilde{K_{\mu }^{0}%
}}$ and $Z_{\mu }$ respectively. As can be seen in Eq. (\ref{32c}), the
vector $\phi _{2}^{0}$ is not orthogonal to $\phi _{3}^{0}.$
Orthogonalizing, we get 
\begin{equation}
\phi _{2}^{0}=N_{2}^{0}\left( \nu _{\chi }\zeta _{\chi }-\frac{\nu _{\eta
}^{2}\nu _{\rho ^{2}}}{\nu _{\rho ^{2}}^{2}+\nu _{\eta }^{2}}\zeta _{\rho
^{2}}-\frac{\nu _{\rho ^{2}}^{2}\nu _{\eta }}{\nu _{\rho ^{2}}^{2}+\nu
_{\eta }^{2}}\zeta _{\eta }\right) .  \label{32f}
\end{equation}%
which correspond to the longitudinal component of $Z_{\mu }^{\prime }$. The
other two eigenvalues different from zero in Eq. (\ref{50}), correspond to
two massive neutral Higgs bosons, whose combination with respect to the
basis of the matrix (\ref{imaginario}) are the vectors $\phi _{4}^{0}$ and $%
\phi _{5}^{0}$ which we rename as $h_{1}^{0}$ and $h_{2}^{0}$ respectively.

\begin{eqnarray}
h_{1}^{0} &=&N_{4}^{0}\left( \nu _{\eta }\nu _{\rho ^{2}}\zeta _{\chi }+\nu
_{\chi }\nu _{\eta }\zeta _{\rho ^{2}}+\nu _{\chi }\nu _{\rho ^{2}}\zeta
_{\eta }\right) ,  \notag \\
h_{2}^{0} &=&N_{5}^{0}\left( \nu _{\chi }\zeta _{\rho ^{3}}-\nu _{\rho
^{2}}\zeta _{\chi ^{0}}\right) .  \label{32i}
\end{eqnarray}

If additionally, we implement the approximations of Eqs. (\ref{27}) and (\ref%
{28}) in the results obtained, the particle spectrum can be written as
indicated in table \ref{tab:doce}.

\begin{table}[!tbh]
\begin{center}
$ 
\begin{tabular}{||c||c||c||}
\hline\hline
Imaginary scalars & Square masses & Features \\ \hline\hline
$\phi _{1}^{0}\simeq \zeta _{\chi ^{0}},$ & $M_{\phi _{1}^{0}}^{2}=0$ & $%
\begin{tabular}{l}
Goldstone \\ 
asociado a $\overline{\widetilde{K_{\mu }^{0}}}$%
\end{tabular}
$ \\ \hline\hline
$\phi _{2}^{0}\simeq \zeta _{\chi }$ & $M_{\phi _{2}^{0}}^{2}=0$ & $%
\begin{tabular}{l}
Goldstone \\ 
asociado a $Z_{\mu }^{\prime }$%
\end{tabular}
$ \\ \hline\hline
$\phi _{3}^{0}=-S_{\theta }\zeta _{\rho ^{2}}+C_{\theta }\zeta _{\eta }$ & $%
M_{\phi _{3}^{0}}^{2}=0$ & 
\begin{tabular}{l}
Goldstone \\ 
asociado a $Z_{\mu }$%
\end{tabular}
\\ \hline\hline
$h_{1}^{0}\simeq C_{\theta }\zeta _{\rho ^{2}}+S_{\theta }\zeta _{\eta }$ & $%
M_{h_{1}^{0}}^{2}\simeq -\frac{2f\nu _{\chi }}{\nu _{\eta }\nu _{\rho ^{2}}}%
(\nu _{\eta }^{2}+\nu _{\rho ^{2}}^{2})$ & Higgs \\ \hline\hline
$h_{2}^{0}\simeq \zeta _{\rho ^{3}}$ & $M_{h_{2}^{0}}^{2}\simeq -\frac{2f\nu
_{\eta }\nu _{\chi }}{\nu _{\rho ^{2}}}+2(\lambda _{8}-2\lambda _{13})\nu
_{\chi }^{2}$ & Higgs \\ \hline\hline
\end{tabular}
$%
\end{center}
\caption{\textit{Physical spectrum of the pseudoscalar fields, for three
Higgs triplets and $\protect\beta =1/\protect\sqrt{3}$.}}
\label{tab:doce}
\end{table}

\noindent The fields $\phi _{1}^{0},$ $\phi _{2}^{0}$ and $\phi _{3}^{0}$
are the three neutral and massless would be Goldstone bosons, that dissapear
from the particle spectrum to give masses to the three neutral gauge bosons $%
\overline{\widetilde{K_{\mu }^{0}}}$, $Z_{\mu }^{\prime }$ and $Z_{\mu }$,
respectively; while $h_{1}^{0}$ and $h_{2}^{0}$ appear in the spectrum as
two neutral Higgs bosons with the masses indicated in table \ref{tab:doce}.
It is taken $f<0$ to define the mass of $h_{1}^{0}$ real and positive.

\subsubsection{Diagonalization of the real sector}

Finally, we take the matrix $M_{\xi \xi }^{2}$ in (\ref{real}) with null
determinant too. The eigenvalues obtained are not simple. So we use the
approximations of the Eqs. (\ref{27}) and (\ref{28}), obtaining the matrix
given in appendix \ref{ap:diagre} Eq. (\ref{53}), which decouples in a null
eigenvalue (would-be Goldstone boson) and the two submatrices $M_{\xi _{\rho
^{2}}\xi _{\eta }}^{2}$ and $M_{\xi _{\chi }\xi _{\rho ^{3}}}^{2}$ indicated
in (\ref{54}). In this manner, it is defined 
\begin{equation}
\phi _{4}^{0}\approx \xi _{\chi ^{0}},  \label{33}
\end{equation}%
when comparing with $\phi _{4}^{0}$ of Appendix \ref{ap:acoplesGg} Eq. (\ref%
{B10}) it is deduced a coupling to the gauge boson $\widetilde{K_{\mu }^{0}}$%
. The submatrix $M_{\xi _{\rho ^{2}}\xi _{\eta }}^{2}$ contains an
eigenvalue zero. However, it does not correspond to a Goldstone boson. Such
eigenvalue arose from the effect of the quadratic approximation in (\ref{53}%
). To extract information for this eigenvalue, we take all the terms of the
sector $M_{\xi _{\rho ^{2}}\xi _{\eta }}^{2}$ from $M_{\xi \xi }^{2},$
obtaining an exact submatrix written in (\ref{55}). Based on this we get the
eigenvalues $\left( P_{2},P_{3}\right) $ associated to the matrix $M_{\xi
_{\rho ^{2}}\xi _{\eta }}^{2}$ of (\ref{55}) and $\left( P_{4},P_{5}\right) $
for the matrix $M_{\xi _{\chi }\xi _{\rho ^{3}}}^{2}$ of (\ref{54}): 
\begin{eqnarray}
P_{2} &=&\frac{1}{2}\left[ \mathcal{M}_{22}+\mathcal{M}_{11}+\sqrt{\left( 
\mathcal{M}_{11}-\mathcal{M}_{22}\right) ^{2}+4\mathcal{M}_{12}^{2}}\right]
\equiv \frac{1}{2}\left[ \mathcal{M}_{D}\mathcal{+}\sqrt{\mathcal{M}_{C}}%
\right] ,  \notag \\
P_{3} &=&\frac{1}{2}\left[ \mathcal{M}_{22}+\mathcal{M}_{11}-\sqrt{\left( 
\mathcal{M}_{11}-\mathcal{M}_{22}\right) ^{2}+4\mathcal{M}_{12}^{2}}\right]
\equiv \frac{1}{2}\left[ \mathcal{M}_{D}\mathcal{-}\sqrt{\mathcal{M}_{C}}%
\right] ,  \notag \\
P_{4} &=&\left[ \Lambda _{+}+\sqrt{\Lambda _{-}^{2}+16\lambda _{10}^{2}}%
\right] \nu _{\chi }^{2},  \notag \\
P_{5} &=&\left[ \Lambda _{+}-\sqrt{\Lambda _{-}^{2}+16\lambda _{10}^{2}}%
\right] \nu _{\chi }^{2},  \label{36 a}
\end{eqnarray}%
with 
\begin{equation}
\Lambda _{\pm }\equiv 4\lambda _{1}\pm \left( \lambda _{8}+2\lambda _{13}-%
\frac{f\nu _{\eta }}{\nu _{\chi }\nu _{\rho ^{2}}}\right) .  \label{36 b}
\end{equation}%
and $\mathcal{M}_{ij}$ denote the elements of the matrix $M_{\xi _{\rho
^{2}}\xi _{\eta }}^{2}$. The non-normalized eigenvectors are written
respectively as

\begin{eqnarray}
h_{3}^{0} &=&S_{\alpha }\xi _{\rho ^{2}}+C_{\alpha }\xi _{\eta },  \notag \\
\quad h_{4}^{0} &=&C_{\alpha }\xi _{\rho ^{2}}-S_{\alpha }\xi _{\eta }, 
\notag \\
\quad h_{5}^{0} &\approx &\left( \Lambda _{-}+\sqrt{\Lambda
_{-}^{2}+16\lambda _{10}^{2}}\right) \xi _{\chi }+4\lambda _{10}\xi _{\rho
^{3}},  \notag \\
\quad h_{6}^{0} &\approx &\left( \Lambda _{-}-\sqrt{\Lambda
_{-}^{2}+16\lambda _{10}^{2}}\right) \xi _{\chi }+4\lambda _{10}\xi _{\rho
^{3}},  \label{37}
\end{eqnarray}%
where $S_{\alpha }\equiv \sin \alpha $, $C_{\alpha }\equiv \cos \alpha $.
Being $\alpha \ $a new mixing angle defined as: 
\begin{eqnarray}
\sin 2\alpha &=&\frac{2\mathcal{M}_{12}}{\sqrt{\left( \mathcal{M}_{11}-%
\mathcal{M}_{22}\right) ^{2}+4\mathcal{M}_{12}^{2}}}  \notag \\
\cos 2\alpha &=&\frac{\mathcal{M}_{22}-\mathcal{M}_{11}}{\sqrt{\left( 
\mathcal{M}_{11}-\mathcal{M}_{22}\right) ^{2}+4\mathcal{M}_{12}^{2}}}.
\label{newangle}
\end{eqnarray}%
the equations above define finally the physical spectrum of the scalar real
fields, as indicated in table \ref{tab:trece}.

\begin{table}[!tbh]
\begin{center}
$ 
\begin{tabular}{||c||c||c||}
\hline\hline
Scalars & Square masses & Features \\ \hline\hline
$\phi _4^0\simeq \xi _{\chi ^0}$ & $M_{\phi _4^0}^2=0$ & 
\begin{tabular}{l}
Goldstone \\ 
of $\widetilde{K_\mu ^0}$%
\end{tabular}
\\ \hline\hline
$h_3^0\simeq S_\alpha \xi _{\rho ^2}+C_\alpha \xi _\eta $ & $%
M_{h_3^0}^2=\frac 12\left[ \mathcal{M}_D+\sqrt{\mathcal{M}_C}\right] ,$ & 
Higgs \\ \hline\hline
$h_4^0\simeq C_\alpha \xi _{\rho ^2}-S_\alpha \xi _\eta $ & $%
M_{h_4^0}^2=\frac 12\left[ \mathcal{M}_D-\sqrt{\mathcal{M}_C}\right] $ & 
Higgs \\ \hline\hline
$h_5^0\simeq \Lambda _1\xi _\chi +4\lambda _{10}\xi _{\rho ^3}$ & $%
M_{h_5^0}^2\simeq \Lambda _1\nu _\chi ^2$ & Higgs \\ \hline\hline
$h_6^0\simeq \Lambda _2\xi _\chi +4\lambda _{10}\xi _{\rho ^3}$ & $%
M_{h_6^0}^2\simeq \Lambda _2\nu _\chi ^2$ & Higgs \\ \hline\hline
\end{tabular}
$%
\end{center}
\caption{\textit{Physical spectrum of masses for the real scalar fields,
with three Higgs triplets and with $\protect\beta =1/\protect\sqrt{3}$. We
use the definitions of Eqs. (\protect\ref{basicdef}).}}
\label{tab:trece}
\end{table}

$\phi _{4}^{0}$ is a neutral would-be Goldstone boson associated to the
gauge field $\widetilde{K_{\mu }^{0}}$ and $h_{3}^{0}$, $h_{4}^{0},$ $%
h_{5}^{0}$, $h_{6}^{0}$ are four neutral Higgs bosons with masses given in
table \ref{tab:trece}; with the condition $f<0$ in order for $h_{4}^{0}$ to
acquire a positive defined mass.

Finally, we are able to summarize the particle spectrum of the scalar sector
and the would be Goldstone bosons in table \ref{tab:catorce}. Taking into
account the following definitions

\begin{eqnarray}
\Lambda _{1,2} &\equiv &\left( \Lambda _{-}\pm \sqrt{\Lambda
_{-}^{2}+16\lambda _{10}^{2}}\right)  \notag \\
\Lambda _{\pm } &=&4\lambda _{1}\pm \left( \lambda _{8}+2\lambda _{13}-\frac{%
f\nu _{\eta }}{\nu _{\chi }\nu _{\rho ^{2}}}\right)  \notag \\
\nu _{\chi } &\gg &\nu _{\rho ^{2}},\nu _{\rho ^{3}},\nu _{\eta }\ \ \ ;\ \
\ \left\vert f\right\vert \approx \nu _{\chi };\qquad f<0,  \notag \\
S_{\theta } &\equiv &\frac{\nu _{\rho ^{2}}}{\sqrt{\nu _{\rho ^{2}}^{2}+\nu
_{\eta }^{2}}},\quad C_{\theta }\equiv \frac{\nu _{\eta }}{\sqrt{\nu _{\rho
^{2}}^{2}+\nu _{\eta }^{2}}},  \notag \\
\sin 2\alpha &\equiv &\frac{2\mathcal{M}_{12}}{\sqrt{\mathcal{M}_{C}}}\ \ ,\
\ \cos 2\alpha \equiv \frac{\mathcal{M}_{22}-\mathcal{M}_{11}}{\sqrt{%
\mathcal{M}_{C}}},  \notag \\
\mathcal{M}_{D} &\equiv &\mathcal{M}_{11}+\mathcal{M}_{22}\text{\ ;\ }%
\mathcal{M}_{C}=\left( \mathcal{M}_{11}-\mathcal{M}_{22}\right) ^{2}+4%
\mathcal{M}_{12}^{2}  \notag \\
\mathcal{M}_{ij} &=&\left( M_{\xi _{\rho ^{2}}\xi _{\eta }}^{2}\right) _{ij}
\label{basicdef}
\end{eqnarray}

We see then that the scalar spectrum consists of eight would be Goldstone
bosons (four charged and four neutral ones) plus ten physical Higgs bosons
(four charged and six neutral ones).

\begin{table}[!tbh]
\begin{center}
$ 
\begin{tabular}{||l||l||l||}
\hline\hline
Scalars & Square masses & Feature \\ \hline\hline
$\phi _1^{\pm }=-S_\theta \rho ^{\pm }+C_\theta \eta ^{2\pm }$ & $M_{\phi
_1^{\pm }}^2=0$ & Goldstone \\ \hline\hline
$\phi _2^{\pm }\simeq \chi ^{\pm }$ & $M_{\phi _2^{\pm }}^2=0$ & Goldstone
\\ \hline\hline
$\phi _1^0\simeq \zeta _{\chi ^0}$ & $M_{\phi _1^0}^2=0$ & Goldstone \\ 
\hline\hline
$\phi _2^0\simeq \zeta _\chi $ & $M_{\phi _2^0}^2=0$ & Goldstone \\ 
\hline\hline
$\phi _3^0=-S_\theta \zeta _{\rho ^2}+C_\theta \zeta _\eta $ & $M_{\phi
_3^0}^2=0$ & Goldstone \\ \hline\hline
$\phi _4^0\simeq \xi _{\chi ^0}$ & $M_{\phi _4^0}^2=0$ & Goldstone \\ 
\hline\hline
$h_1^{\pm }\simeq \eta ^{3\pm }$ & $M_{h_1^{\pm }}^2\simeq \lambda _7\nu
_\chi ^2-f\nu _\chi \frac{\nu _{\rho ^2}}{\nu _\eta }$ & Higgs \\ 
\hline\hline
$h_2^{\pm }\simeq C_\theta \rho ^{\pm }+S_\theta \eta ^{2\pm }$ & $%
M_{h_2^{\pm }}^2\simeq -f\nu _\chi \left( \frac{\nu _\eta }{\nu _{\rho ^2}}+%
\frac{\nu _{\rho ^2}}{\nu _\eta }\right) $ & Higgs \\ \hline\hline
$h_1^0\simeq C_\theta \zeta _{\rho ^2}+ S_\theta \zeta _\eta $ & $%
M_{h_1^0}^2\simeq -\frac{2f\nu _\chi }{\nu _\eta \nu _{\rho ^2}}\left( \nu
_\eta ^2+\nu _{\rho ^2}^2\right) $ & Higgs \\ \hline\hline
$h_2^0\simeq \zeta _{\rho ^3}$ & $M_{h_2^0}^2\simeq -\frac{2f\nu _\eta \nu
_\chi }{\nu _{\rho ^2}}+2(\lambda _8-2\lambda _{13})\nu _\chi ^2$ & Higgs \\ 
\hline\hline
$h_3^0\simeq S_\alpha \xi _{\rho ^2}+C_\alpha \xi _\eta $ & $%
M_{h_3^0}^2\approx \nu_\eta\nu_{\rho^2}$ & Higgs \\ \hline\hline
$h_4^0\simeq C_\alpha \xi _{\rho ^2}-S_\alpha \xi _\eta $ & $%
M_{h_4^0}^2\approx f \nu_\chi$ & Higgs \\ \hline\hline
$h_5^0\simeq \Lambda _1\xi _\chi +4\lambda _{10}\xi _{\rho ^3}$ & $%
M_{h_5^0}^2\simeq \Lambda _1\nu _\chi ^2$ & Higgs \\ \hline\hline
$h_6^0\simeq \Lambda _2\xi _\chi +4\lambda _{10}\xi _{\rho ^3}$ & $%
M_{h_6^0}^2\simeq \Lambda _2\nu _\chi ^2$ & Higgs \\ \hline\hline
\end{tabular}
$%
\end{center}
\caption{\textit{Physical scalar spectrum after both transitions, for three
Higgs triplets with $\protect\beta =1/\protect\sqrt{3}$. Taking into account
the definitions in Eqs. (\protect\ref{basicdef}).}}
\label{tab:catorce}
\end{table}

\section{Trilinear Higgs-Gauge Bosons terms for $\protect\beta =1/\protect%
\sqrt{3}$}

The term \textbf{2 }in Eq. (\ref{B1}) contains cubic couplings between
scalar and gauge bosons. We consider these couplings for the SM gauge
bosons: $W_{\mu }^{\pm },$ $Z_{\mu }$ and $A_{\mu },$ with masses given by 
\cite{fourteen}:

\begin{eqnarray}
M_{W^{\pm }} &=&\frac g{\sqrt{2}}\sqrt{\nu _{\rho ^2}^2+\nu _\eta ^2}  \notag
\\
M_Z &=&\frac g{\sqrt{2}C_W}\sqrt{\nu _{\rho ^2}^2+\nu _\eta ^2}.
\label{vectormass}
\end{eqnarray}

In Eq. (\ref{B1}) there are not cubic couplings of electroweak bosons with
the scalar boson $\chi $ (that acquire VEV in the 3$^{\underline{rd}}$
component), as can be seen in Eq. (\ref{B5}), where the SM gauge bosons are
in the $2\times 2$ superior components, while in the third component ($3^{%
\underline{rd}}$ row and column) are the exotic gauge bosons which get their
masses through $v_{\chi }$. Using Eqs. (\ref{B5}) and (\ref{B6}) for $\Phi
=\rho $ and $\eta $ in Eq. (\ref{B1}), it is found 
\begin{eqnarray}
L_{trilinear}^{SM} &=&g^{2}\left( \nu _{\rho ^{2}}\xi _{\rho ^{2}}+\nu
_{\eta }\xi _{\eta }\right) W^{\mu -}W_{\mu }^{+}+2a_{4}^{2}\left( \nu
_{\rho ^{2}}\xi _{\rho ^{2}}+\nu _{\eta }\xi _{\eta }\right) Z^{\mu }Z_{\mu }
\notag \\
&+&\frac{g}{\sqrt{2}}a_{1}\left( -\nu _{\rho ^{2}}\rho ^{-}+\nu _{\eta }\eta
_{2}^{-}\right) A^{\mu }W_{\mu }^{+}+h.c  \notag \\
&+&\frac{g}{\sqrt{2}}(a_{2}+a_{4})\left( -\nu _{\rho ^{2}}\rho ^{-}+\nu
_{\eta }\eta _{2}^{-}\right) Z^{\mu }W_{\mu }^{+}  \notag \\
&-&\frac{ig}{\sqrt{2}}\partial ^{\mu }\rho ^{-}W_{\mu }^{+}\xi _{\rho ^{2}}-%
\frac{ig}{\sqrt{2}}\partial ^{\mu }\xi _{\rho ^{2}}W_{\mu }^{-}\rho ^{+} 
\notag \\
&-&\frac{ig}{\sqrt{2}}\partial ^{\mu }\eta _{2}^{+}W_{\mu }^{-}\xi _{\eta }-%
\frac{ig}{\sqrt{2}}\partial ^{\mu }\xi _{\eta }W_{\mu }^{+}\eta _{2}^{-}+h.c,
\label{cubic}
\end{eqnarray}%
where $a_{1},a_{2}$ and $a_{4}$ are defined in Eq. (\ref{B6}). The terms in
Eq. (\ref{cubic}) are in the weak basis, which are related to the physical
basis through table \ref{tab:catorce}. In the physical basis, and after some
algebraic manipulation using Eq. (\ref{B6}) in Eq. (\ref{cubic}), it is
finally found 
\begin{eqnarray}
L_{trilinear}^{SM} &=&gM_{W^{\pm }}\cos \left( \theta -\alpha \right) W^{\mu
-}W_{\mu }^{+}h_{3}^{0}+gM_{W^{\pm }}\sin \left( \theta -\alpha \right)
W^{\mu -}W_{\mu }^{+}h_{4}^{0}  \notag \\
&+&\frac{g}{2C_{W}}M_{Z}\cos \left( \theta -\alpha \right) Z^{\mu }Z_{\mu
}h_{3}^{0}+\frac{g}{2C_{W}}M_{Z}\sin \left( \theta -\alpha \right) Z^{\mu
}Z_{\mu }h_{4}^{0}  \notag \\
&-&\left( eM_{W^{\pm }}A^{\mu }W_{\mu }^{+}\phi
_{1}^{-}+gM_{Z}S_{W}^{2}Z^{\mu }W_{\mu }^{+}\phi _{1}^{-}+h.c\right)  \notag
\\
&-&\frac{g}{2}\left( p-k\right) ^{\mu }\cos \left( \theta -\alpha \right)
W_{\mu }^{+}h_{2}^{-}h_{4}^{0}+\frac{g}{2}\left( p-q\right) ^{\mu }\sin
\left( \theta -\alpha \right) W_{\mu }^{+}h_{2}^{-}h_{3}^{0}  \notag \\
&-&\frac{ig}{2}\left( p-r\right) ^{\mu }W_{\mu }^{+}h_{2}^{-}h_{1}^{0}+h.c,
\end{eqnarray}%
where the electric charge has been defined as $e=gS_{W},$ and $\theta $ and $%
\alpha $ are given by Eqs. (\ref{32}) and (\ref{newangle}). It worths to
note that these vertices are the same as the ones obtained in the Standard
Model with two Higgs doublets (2HDM). In the notation of Ref. \cite{Hunter},
the Higgs bosons of the 2HDM have the following correspondence with the
Higgs notation shown here%
\begin{equation*}
h_{3}^{0}\rightarrow h^{0}\text{,}\ \ h_{4}^{0}\rightarrow H^{0}\text{,}\ \
h_{1}^{0}\rightarrow A^{0}\text{,\ \ }h_{2}^{\pm }\rightarrow H^{\pm },\ \
\phi _{1}^{\pm }\rightarrow G_{W}^{\pm }\text{,\ \ }\phi _{3}^{0}\rightarrow
G_{Z}^{0}
\end{equation*}%
The couplings of the charged Higgs bosons to the vertex $A_{\mu }W^{\mu }$
vanish because of the conservation of electromagnetic charge and only the
couplings to the would-be Goldstone bosons can exist since they are not
physical fields.

In fact, we can realize that the aproximation $f\nu _{x}>>\nu _{\rho
^{2}}^{2}$,$\nu _{\eta }^{2}$ implies that the boson $h_{4}^{0}$ (very
massive Higgs) decouple from the vertices, and that $h_{3}^{0}$ couple in
the same way as the SM Higgs boson, without any other constraint. In this
limit, it is obtained that 
\begin{equation}
\tan 2\alpha \approx \frac{2\nu _{\rho ^{2}}\nu _{\eta }}{\nu _{\eta
}^{2}-\nu _{\rho ^{2}}}
\end{equation}%
i.e. $\theta \approx \alpha $, and the $L_{trilinear}^{SM}$ is given by 
\begin{eqnarray}
L_{trilinear}^{SM} &=&gM_{W^{\pm }}W^{\mu -}W_{\mu }^{+}h_{3}^{0}+\frac{g}{%
2C_{W}}M_{Z}Z^{\mu }Z_{\mu }h_{3}^{0}  \notag \\
&-&\left( eM_{W^{\pm }}A^{\mu }W_{\mu }^{+}\phi
_{1}^{-}-gM_{Z}S_{W}^{2}Z^{\mu }W_{\mu }^{+}\phi _{1}^{-}+h.c\right)  \notag
\\
&-&\left( \frac{g}{2}\left( p-k\right) ^{\mu }W_{\mu }^{+}h_{2}^{-}h_{4}^{0}+%
\frac{ig}{2}\left( p-r\right) ^{\mu }W_{\mu
}^{+}h_{2}^{-}h_{1}^{0}+h.c.\right) .  \label{trilinear limit}
\end{eqnarray}

Where Eq. (\ref{trilinear limit}) shows explicitly the fact that the
couplings of $h_{3}^{0}$ are SM-like when the limit $f\nu _{x}>>\nu _{\rho
^{2}}^{2}$,$\nu _{\eta }^{2}$; is taken.

\section{Low energy limit for $\protect\beta =1/\protect\sqrt{3}$\label{low
energy}}

We shall discuss briefly some important properties of the low energy limit
of this model. As we see in table \ref{tab:catorce}, the Higgs scalar masses
depend on some parameters of the Higgs potential, the vacuum expectation
value $\nu _{\chi }$ at the higher scale, and the vacuum expectation values
at the electroweak scale ($\nu _{\eta }$, $\nu _{\rho ^{2}}$) as it must be.
Notwithstanding, there is a crucial parameter that determines the scale in
which the Higgs bosons lie, i.e. the trilinear coupling constant $f$ defined
in Eq. (\ref{21a}). For arbitrary values of it (or more precisely for
arbitrary values of $f\nu _{\chi }$) some of the Higgs bosons would lie on a
new scale different from the breakdown ones. If we argue a naturalness
criterium, we could assume that $f\nu _{\chi }$ should lie either in the
electroweak or $SU\left( 3\right) _{L}\otimes U\left( 1\right) _{X}$
breaking scale. Since these 331 models break to a $SU\left( 2\right)
_{L}\otimes U\left( 1\right) _{Y}\ $two\ Higgs doublet model; the value
assumed for $f\nu _{\chi }$ determines the scales for the $SU\left( 2\right)
_{L}\otimes U\left( 1\right) _{Y}$ Higgs bosons which can be identified as
the $h_{2}^{\pm }$,$\ h_{1}^{0},h_{3}^{0}$ and $h_{4}^{0}$ fields of table %
\ref{tab:catorce}, where $\phi _{1}^{\pm }\ $and $\phi _{3}^{0}$ are the
corresponding would-be Goldstone bosons. We also see in table \ref%
{tab:catorce} that $h_{3}^{0}$ is the only Higgs field that depends
exclusively on the vacuum expectation values at the electroweak scale, so it
corresponds to the SM Higgs. The scale of the other $SU\left( 2\right)
_{L}\otimes U\left( 1\right) _{Y}$ Higgs bosons depend on the value assumed
by $f\nu _{\chi }$. If $f\nu _{\chi }\sim O\left( \nu _{\chi }^{2}\right) $
the remaining $SU\left( 2\right) _{L}\otimes U\left( 1\right) _{Y}$ Higgs
bosons belong to the higher breaking scale and for sufficiently large values
of $\nu _{\chi }^{2}$, the $h_{2}^{\pm }$,$\ h_{1}^{0}$ and $h_{4}^{0}$
fields become heavy modes, in addition by taking $f\nu _{\chi }$ large
enough we obtain automatically that the two mixing angles of the Higgs
sector in the remnant two Higgs doublet model become equal, from which the
decoupling limit at the electroweak scale to the minimal SM follows naturally%
\cite{decoupling}. Otherwise, if we assume $f\nu _{\chi }\sim O\left( \nu
_{EW}^{2}\right) $ \ the diagonalization process for the scalar sector would
be much more complicated, but a naive analysis shows that we would get a
spectrum of $SU\left( 2\right) _{L}\otimes U\left( 1\right) _{Y}$ Higgs
bosons lying roughly on the EW scale, obtaining a non-decoupling two Higgs
doublet model.

%%%%%%%%%%%%%%%%%%%%%%%%%%%%%%%%%%%%%
%%%%%%%%%%%%%%%%%%%%%%%%%%%%%%%%%%%%%%

\section{Conclusions\label{conclusions}}

For the 331 models, the set of equations arising from anomalies predicts
that the number of triplets and antitriplets must be equal in order to
cancel anomalies. These equations lead to an infinite set of possible
solutions. If we impose for the model not to have exotic electromagnetic
charges, the number of solutions is reduced to only two. On the other hand,
the quantum numbers may be restricted by demanding gauge invariance for the
Yukawa couplings of the scalar fields to fermions (classical constraints) as
well as from the cancelation of fermionic anomalies (quantum constraints),
obtaining models with ordinary and exotic charges. Letting free the
parameters $X\ $and $\beta $ that define the electromagnetic charge, we get
all possible vacuum alignments in both symmetry breakings. When we choice
the VEV's according to the breakdown scheme $331\rightarrow 321\rightarrow 1$
the values for $X$ and $\beta $ are found, obtaining the models already
studied in the literature plus other ones with exotic charges. Moreover, it
is found that one scalar triplet is necessary for the first breaking and two
scalar triplets for the second one, the two triplets for the second breaking
contain two doublets of $SU(2)_{L}$ for the second breaking that give masses
to the up and down quarks respectively. In some cases it is necessary to
introduce an additional scalar sextet in order to endow the neutrinos with
masses.

Furthermore, we present the most general Higgs potentials for the models
with $\beta =\pm \sqrt{3}$, $\beta =\pm 1/\sqrt{3}$ and $\beta $ arbitrary
with three scalar triplets and one scalar sextet. In particular, the case of 
$\beta =1/\sqrt{3}$ is analized, getting the scalar spectrum in the $%
R_{\zeta }-$gauge; such spectrum consists of ten massive Higgs bosons as
well as eight would be Goldstone bosons necessary to break the eight
generators of the SSB scheme demanded.

Besides, in the framework of the $\beta =1/\sqrt{3}$ model, we calculate the
trilinear couplings of the Higgs bosons to the SM gauge bosons, finding that
such couplings are identical to the ones of the standard two Higgs doublet
model (2HDM). It worths to say that $A_{\mu } W_{\mu }^{\pm }$ are not
coupled to the charged Higgs bosons in consistence with electromagnetic
charge conservation.

As for the low energy limit ($\nu _{x}^{2}>\!>\nu _{\rho ^{2}}^{2}$, $\nu
_{\eta }^{2}$), the Higgs potential of the 331 model with $\beta =1/\sqrt{3}$
is also reduced to the Higgs potential of the 2HDM in this limit. Thus, this
model ends up in a $SU\left( 2\right) _{L}\otimes U\left( 1\right) _{Y}$ two
Higgs doublet model after the first symmetry breaking. We see that the
behavior of the scalar sector of these models at low energies is strongly
determined by the trilinear coupling $f$ of the Higgs potential. For
instance, by setting $f\nu _{\chi }\sim O\left( \nu _{EW}^{2}\right) $ all
the $SU\left( 2\right) _{L}\otimes U\left( 1\right) _{Y}$ Higgs bosons
belong to the electroweak scale, obtaining a non-decoupling two Higgs
doublet model. By contrast, the assumption $f\nu _{\chi }\sim O\left( \nu
_{\chi }^{2}\right) $ leaves one of these Higgs bosons at the electroweak
scale (the SM Higgs), and the other four ones would lie on the scale $%
O\left( \nu _{\chi }\right) $ of the first transition; additionally, if the
scale $O\left( \nu _{\chi }\right) $ is sufficiently large, we obtain that
the mixing angles of the Higgs potential are equal, from which the couplings
of the light Higgs boson to fermions and gauge fields become identical to
the SM couplings. Therefore, the natural assumption $f\nu _{\chi }\sim
O\left( \nu _{\chi }^{2}\right) >\!>\nu _{\rho ^{2}}^{2}$, $\nu _{\eta }^{2}$
leads automatically to the minimal SM at low energies.

\section*{Acknowledgments}

We thank to \emph{Fundaci\'{o}n Banco de la Rep\'{u}blica} for its financial
support.

\setcounter{equation}{0} \appendix \numberwithin{equation}{section} 

\section{Scalar masses\label{ap:S masses}}

\subsubsection*{i)\ Imaginary sector}

The mass matrix $M_{\zeta \zeta }^{2}$ in the basis $\zeta _{\chi },\zeta
_{\rho ^{2}},\zeta _{\rho ^{3}},\zeta _{\eta },\zeta _{\chi ^{0}}$ is given
by 
\begin{eqnarray}
M_{\zeta \zeta }^{2}(1,1) &=&(2\lambda _{8}-4\lambda _{13})\nu _{\rho
^{3}}^{2}-f(\frac{2\nu _{\eta }\nu _{\rho ^{2}}}{\nu _{\chi }}+\frac{2\nu
_{\eta }\nu _{\rho ^{3}}^{2}}{\nu _{\chi }\nu _{\rho ^{2}}})  \notag \\
M_{\zeta \zeta }^{2}(1,2)=M_{\zeta \zeta }^{2}(2,1) &=&-2f\nu _{\eta } 
\notag \\
M_{\zeta \zeta }^{2}(1,3)=M_{\zeta \zeta }^{2}(3,1) &=&(2\lambda
_{8}+4\lambda _{13})\nu _{chi}\nu _{\rho ^{3}}+2f\frac{\nu _{\rho ^{3}}\nu
_{eta}}{\nu _{\rho ^{2}}}  \notag \\
M_{\zeta \zeta }^{2}(1,4)=M_{\zeta \zeta }^{2}(4,1) &=&-2f\nu _{\rho ^{2}} 
\notag \\
M_{\zeta \zeta }^{2}(1,5)=M_{\zeta \zeta }^{2}(5,1) &=&(2\lambda
_{8}-4\lambda _{13})\nu _{\rho ^{2}}\nu _{\rho ^{3}}  \notag \\
M_{\zeta \zeta }^{2}(2,2) &=&-2f\frac{\nu _{\eta }\nu _{\chi }}{\nu _{\rho
^{2}}}  \notag \\
M_{\zeta \zeta }^{2}(2,3)=M_{\zeta \zeta }^{2}(3,2) &=&0  \notag \\
M_{\zeta \zeta }^{2}(2,4)=M_{\zeta \zeta }^{2}(4,2) &=&-2f\nu _{\chi } 
\notag \\
M_{\zeta \zeta }^{2}(2,5)=M_{\zeta \zeta }^{2}(5,2) &=&2f\frac{\nu _{\eta
}\nu _{\rho ^{3}}}{\nu _{\rho ^{2}}}  \notag \\
M_{\zeta \zeta }^{2}(3,3) &=&\left( 2\lambda _{8}-4\lambda _{13}\right) \nu
_{\chi }^{2}-2f\frac{\nu _{\chi }\nu _{\eta }}{\nu _{\rho ^{2}}}  \notag \\
M_{\zeta \zeta }^{2}(3,4)=M_{\zeta \zeta }^{2}(4,3) &=&0  \notag \\
M_{\zeta \zeta }^{2}(3,5)=M_{\zeta \zeta }^{2}(5,3) &=&\left( 2\lambda
_{8}-4\lambda _{13}\right) \nu _{\chi }^{2}-2f\frac{\nu _{\chi }\nu _{\eta }%
}{\nu _{\rho ^{2}}}  \notag \\
M_{\zeta \zeta }^{2}(4,4) &=&-2f\frac{\nu _{\chi }\nu _{\rho ^{2}}}{\nu
_{\eta }}  \notag \\
M_{\zeta \zeta }^{2}(4,5)=M_{\zeta \zeta }^{2}(5,4) &=&2f\nu _{\rho ^{3}} 
\notag \\
M_{\zeta \zeta }^{2}(5,5) &=&(2\lambda _{8}-4\lambda _{13})\nu _{\rho
^{2}}^{2}-f\left( \frac{2\nu _{\eta }\nu _{\rho ^{2}}}{\nu _{\chi }}+\frac{%
2\nu _{\eta }\nu _{\rho ^{3}}^{2}}{\nu _{\chi }\nu _{\rho ^{2}}}\right)
\label{imaginario}
\end{eqnarray}

\subsubsection*{ii.) Real sector}

The mass matrix $M_{\xi \xi }^{2}$ in the basis $\xi _{\chi },\xi _{\rho
^{2}},\xi _{\rho ^{3}},\xi _{\eta },\xi _{\chi ^{0}}$ is given by 
\begin{eqnarray}
M_{\xi \xi }^{2}(1,1) &=&8\lambda _{1}\nu _{\chi }^{2}+8\lambda _{10}\nu
_{\chi }\nu _{\rho ^{3}}+\left( 2\lambda _{8}+4\lambda _{13}\right) \nu
_{\rho ^{3}}^{2}  \notag \\
&-&f\left( \frac{2\nu _{\eta }\nu _{\rho ^{2}}}{\nu _{\chi }}+\frac{2\nu
_{\eta }\nu _{\rho ^{3}}^{2}}{\nu _{\chi }\nu _{\rho ^{2}}}\right)  \notag \\
M_{\xi \xi }^{2}(1,2)=M_{\xi \xi }^{2}(2,1) &=&4\lambda _{5}\nu _{\chi }\nu
_{\rho ^{2}}+4\lambda _{11}\nu _{\rho ^{2}}\nu _{\rho ^{3}}+2f\nu _{\eta } 
\notag \\
M_{\xi \xi }^{2}(1,3)=M_{\xi \xi }^{2}(3,1) &=&(4\lambda _{5}+2\lambda
_{8}+4\lambda _{13})\nu _{\chi }\nu _{\rho ^{3}}+4\lambda _{10}\nu _{\chi
}^{2}+4\lambda _{11}\nu _{\rho ^{3}}^{2}+2f\frac{\nu _{\eta }\nu _{\rho ^{3}}%
}{\nu _{\rho ^{2}}}  \notag \\
M_{\xi \xi }^{2}(1,4)=M_{\xi \xi }^{2}(4,1) &=&4\lambda _{4}\nu _{\chi }\nu
_{\eta }+4\lambda _{12}\nu _{\eta }  \notag \\
M_{\xi \xi }^{2}(1,5)=M_{\xi \xi }^{2}(5,1) &=&(2\lambda _{8}+4\lambda
_{13})\nu _{\rho ^{2}}\nu _{\rho ^{3}}+4\lambda _{10}\nu _{\chi }\nu _{\rho
^{2}}  \notag \\
M_{\xi \xi }^{2}(2,2) &=&8\lambda _{2}\nu _{\rho ^{2}}^{2}-2f\frac{\nu
_{\eta }\nu _{\chi }}{\nu _{\rho ^{2}}}  \notag \\
M_{\xi \xi }^{2}(2,3)=M_{\xi \xi }^{2}(3,2) &=&8\lambda _{2}\nu _{\rho
^{2}}\nu _{\rho ^{3}}+4\lambda _{11}\nu _{\chi }\nu _{\rho ^{2}}  \notag \\
M_{\xi \xi }^{2}(2,4)=M_{\xi \xi }^{2}(4,2) &=&4\lambda _{6}\nu _{\eta }\nu
_{\rho ^{2}}+2f\nu _{\chi }  \notag \\
M_{\xi \xi }^{2}(2,5)=M_{\xi \xi }^{2}(5,2) &=&4\lambda _{11}\nu _{\rho
^{2}}^{2}+2f\frac{\nu _{\eta }\nu _{\rho ^{3}}}{\nu _{\rho ^{2}}}  \notag \\
M_{\xi \xi }^{2}(3,3) &=&\left( 2\lambda _{8}+4\lambda _{13}\right) \nu
_{\chi }^{2}+8\lambda _{11}\nu _{\chi }\nu _{\rho ^{3}}+8\lambda _{2}\nu
_{\rho ^{3}}^{2}-2f\frac{\nu _{\chi }\nu _{\eta }}{\nu _{\rho ^{2}}}  \notag
\\
M_{\xi \xi }^{2}(3,4)=M_{\xi \xi }^{2}(4,3) &=&4\lambda _{6}\nu _{\eta }\nu
_{\rho ^{3}}+4\lambda _{12}\nu _{\eta }\nu _{\chi }  \notag \\
M_{\xi \xi }^{2}(3,5)=M_{\xi \xi }^{2}(5,3) &=&(2\lambda _{8}+4\lambda
_{13})\nu _{\chi }\nu _{\rho ^{2}}+4\lambda _{11}\nu _{\rho ^{2}}\nu _{\rho
^{3}}-2f\nu _{\eta }  \notag \\
M_{\xi \xi }^{2}(4,4) &=&8\lambda _{3}\nu _{\eta }^{2}-2f\frac{\nu _{\chi
}\nu _{\rho ^{2}}}{\nu _{\eta }}  \notag \\
M_{\xi \xi }^{2}(4,5)=M_{\xi \xi }^{2}(5,4) &=&4\lambda _{12}\nu _{\eta }\nu
_{\rho ^{2}}-2f\nu _{\rho ^{3}}  \notag \\
M_{\xi \xi }^{2}(5,5) &=&(2\lambda _{8}+4\lambda _{13})\nu _{\rho
^{2}}^{2}-f(\frac{2\nu _{\eta }\nu _{\rho ^{2}}}{\nu _{\chi }}+\frac{2\nu
_{\eta }\nu _{\rho ^{3}}^{2}}{\nu _{\chi }\nu _{\rho ^{2}}})  \label{real}
\end{eqnarray}

\subsubsection*{iii.) Charged sector}

The mass matrix $M_{\phi \pm }^{2}$ in the basis $\chi ^{\pm },\rho ^{\pm
},\eta ^{2\pm },\eta ^{3\pm }$ reads

\begin{eqnarray}
M_{\phi \pm }^2(1,1)&=&\lambda _7\nu _\eta ^2 -f\left( \frac{\nu _\eta \nu
_{\rho ^2}}{\nu _\chi } +\frac{\nu _\eta \nu _{\rho ^3}^2}{\nu _\chi \nu
_{\rho ^2}}\right)  \notag \\
M_{\phi \pm }^2(1,2)=M_{\phi \pm }^2(2,1)&=& \lambda _{14}\nu _\eta ^2+f%
\frac{\nu _\eta \nu _{\rho ^3}}{\nu _{\rho ^2}}  \notag \\
M_{\phi \pm }^2(1,3)=M_{\phi \pm }^2(3,1)&=& \lambda _{14}\nu _\eta \nu
_{\rho ^2} +f\nu _{\rho ^3}  \notag \\
M_{\phi \pm }^2(1,4)=M_{\phi \pm }^2(4,1)&=&\lambda _7\nu _\chi \nu
_\eta+\lambda _{14}\nu _\eta \nu _{\rho ^3} -f\nu _{\rho ^2}  \notag \\
M_{\phi \pm }^2(2,2)&=& \lambda _9\nu _\eta ^2 -f\frac{\nu _\eta \nu _\chi }{%
\nu _{\rho ^2}}  \notag \\
M_{\phi \pm }^2(2,3)=M_{\phi \pm }^2(3,2)&=& \lambda _9\nu _\eta \nu _{\rho
^2} -f\nu _\chi  \notag \\
M_{\phi \pm }^2(2,4)=M_{\phi \pm }^2(4,2)&=&\lambda _9\nu _\eta \nu _{\rho
^3} +\lambda _{14}\nu _\chi \nu _\eta  \notag \\
M_{\phi \pm }^2(3,3)&=&\lambda _9\nu _{\rho ^2}^2-f\frac{\nu _{\rho ^2}\nu
_\chi }{\nu _\eta }  \notag \\
M_{\phi \pm }^2(3,4)=M_{\phi \pm }^2(4,3)&=& \lambda _9\nu _{\rho ^2}\nu
_{\rho ^3} +\lambda _{14}\nu _\chi \nu _{\rho ^2}  \notag \\
M_{\phi \pm }^2(4,4)&=& \lambda _7\nu _\chi ^2 +2\lambda _{14}\nu _\chi \nu
_{\rho ^3} +\lambda _9\nu _{\rho ^3}^2 -f\frac{\nu _\chi \nu _{\rho ^2}}{\nu
_\eta }  \label{cargado}
\end{eqnarray}

The previous matrices are singular, it is associated to the would be
Goldstone bosons of the theory (at least one null eigenvalue). 
\begin{equation}
\det (M_{\xi \xi }^{2})=\det (M_{\zeta \zeta }^{2})=\det (M_{\phi \pm
}^{2})=0.
\end{equation}

\subsection{Diagonalization of the charged sector\label{ap:diagcar}}

The matrix $M_{\phi \pm }^{2}$ in Eq. (\ref{cargado}) presents null
eigenvalues. For the eigenvalue $P_{1}=0$, the equation (\ref{25}) is block
reduced to:

\begin{equation}
\left[ 
\begin{array}{cccc}
1 & 0 & 0 & \frac{\nu _\chi }{\nu _\eta } \\ 
0 & 1 & \frac{\nu _{\rho ^2}}{\nu _\eta } & \frac{\nu _{\rho ^3}}{\nu _\eta }
\\ 
0 & 0 & 0 & 0 \\ 
0 & 0 & 0 & 0%
\end{array}
\right] \left( 
\begin{array}{c}
x_1 \\ 
x_2 \\ 
x_3 \\ 
x_4%
\end{array}
\right) =0,  \label{41}
\end{equation}

\noindent it produces two rows of zeros because of the fact that the zero
eigenvalue is two-fold degenerate, such that only two of the four equations
defined by (\ref{25}) are linearly independent, as can be seen in Eq. (\ref%
{41}). Since there are four variables, we take a particular solution among
the infinite ones, for instance

\begin{equation}
x_3=1;\qquad x_4=0,  \label{42}
\end{equation}

\noindent and when we replace in (\ref{41}), it is found

\begin{equation}
x_{1}=0;\qquad x_{2}=-\frac{\nu _{\rho ^{2}}}{\nu _{\eta }}.  \label{43}
\end{equation}%
obtaining the solution $\phi _{1}^{\pm }$ written in Eq. (\ref{26a}).

To find out the solution of the other degenerate eigenvalue $P_{2}=0,$ we
choose 
\begin{equation}
x_{3}=0;\qquad x_{4}=-1,  \label{44}
\end{equation}%
and from Eq. (\ref{41}), we find 
\begin{equation}
x_{1}=\frac{\nu _{\chi }}{\nu _{\eta }};\qquad x_{2}=\frac{\nu _{\rho ^{3}}}{%
\nu _{\eta }}.  \label{45}
\end{equation}%
obtaining $\phi _{2}^{\pm }$ in Eq. (\ref{26a}).

For the other two eigenvalues we utilize the approximation given by Eqs. (%
\ref{27}). Keeping only quadratic order in $\nu _{\chi }$, it is obtained
from the matrix in Eq. (\ref{cargado}): 
\begin{eqnarray}
&&%
\begin{array}{ccccc}
\quad \mathbf{\chi }^{\pm } &  & \mathbf{\rho }^{\pm } & \mathbf{\eta }%
^{2\pm } & \;\mathbf{\eta }^{3\pm }%
\end{array}
\\
M_{\phi \pm }^{2} &\sim &\left[ 
\begin{array}{cccc}
0\quad & 0\quad & 0\quad & 0 \\ 
0\quad & 0\quad & 0\quad & 0 \\ 
0\quad & 0\quad & 0\quad & 0 \\ 
0\quad & 0\quad & 0\quad & \lambda _{7}\nu _{\chi }^{2}%
\end{array}%
\right] .  \label{47}
\end{eqnarray}

\noindent with only one eigenvalue. To be able to find out the other
eigenvalue, we demand the condition of the Eq. (\ref{28}), and the matrix (%
\ref{cargado}) can be approximated at order $\nu _{\chi }^{2}$, $f\nu _{\chi
}$ to the form

\begin{eqnarray}
&&%
\begin{array}{cccc}
\quad \mathbf{\chi }^{\pm } & \mathbf{\rho }^{\pm }\qquad & \quad \mathbf{%
\eta }^{2\pm }\qquad & \mathbf{\eta }^{3\pm }\qquad \qquad%
\end{array}
\notag \\
M_{\phi \pm }^{2} &\approx &\left[ 
\begin{array}{cccc}
0 & 0 & 0 & 0 \\ 
0 & -f\frac{\nu _{\eta }\nu _{\chi }}{\nu _{\rho ^{2}}} & -f\nu _{\chi } & 0
\\ 
0 & -f\nu _{\chi } & -f\frac{\nu _{\rho ^{2}}\nu _{\chi }}{\nu _{\eta }} & 0
\\ 
0 & 0 & 0 & \lambda _{7}\nu _{\chi }^{2}-f\frac{\nu _{\chi }\nu _{\rho ^{2}}%
}{\nu _{\eta }}%
\end{array}%
\right] ,  \label{48}
\end{eqnarray}%
which is block reduced in the following way 
\begin{eqnarray}
M_{\chi ^{\pm }}^{2} &=&0  \notag \\
&&%
\begin{array}{cc}
\quad \quad \mathbf{\rho }^{\pm } & \quad \quad \mathbf{\eta }^{2\pm }%
\end{array}
\notag \\
M_{\rho \eta ^{2}}^{2} &\approx &\left[ 
\begin{array}{cc}
-f\frac{\nu _{\eta }\nu _{\chi }}{\nu _{\rho ^{2}}} & -f\nu _{\chi } \\ 
-f\nu _{\chi } & -f\frac{\nu _{\rho ^{2}}\nu _{\chi }}{\nu _{\eta }}%
\end{array}%
\right] ,  \notag \\
M_{\eta ^{3\pm }}^{2} &\approx &\lambda _{7}\nu _{\chi }^{2}-f\frac{\nu
_{\chi }\nu _{\rho ^{2}}}{\nu _{\eta }}  \label{49}
\end{eqnarray}%
The eigenvalue$\;P_{3}=\lambda _{7}\nu _{\chi }^{2}-f\nu _{\chi }\frac{\nu
_{\rho ^{2}}}{\nu _{\eta }}$ , is associated to the decoupled component $%
\eta ^{3\pm }$ as it is written in Eq. (\ref{28b}). The square matrix in Eq.
(\ref{49}) contains one of the null degenerate eigenvalues. The other one is
the $P_{4}\;$eigenvalue written in Eq. (\ref{29}), whose eigenvector is
given by Eq. (\ref{31a}).

\subsection{Diagonalization of the imaginary sector\label{ap:diagim}}

As for the matrix in Eq. (\ref{imaginario}), its eigenvalues read

\begin{eqnarray}
P_{1} &=&0,  \notag \\
P_{2} &=&0,  \notag \\
P_{3} &=&0,  \notag \\
P_{4} &=&-\frac{2f}{\nu _{\eta }\nu _{\rho ^{2}}\nu _{\chi }}\left( \nu
_{\chi }^{2}\nu _{\eta }^{2}+\nu _{\chi }^{2}\nu _{\rho ^{2}}^{2}+\nu _{\rho
^{3}}^{2}\nu _{\eta }^{2}+\nu _{\rho ^{2}}^{2}\nu _{\eta }^{2}\right) , 
\notag \\
P_{5} &=&-\frac{2}{\nu _{\rho ^{2}}\nu _{\chi }}\left[ \left( 2\lambda
_{13}-\lambda _{8}\right) \nu _{\rho ^{2}}\nu _{\chi }^{3}+f\nu _{\eta }\nu
_{\chi }^{2}+f\nu _{\eta }\left( \nu _{\rho ^{2}}^{2}+\nu _{\rho
^{3}}^{2}\right) \right.  \notag \\
&&\left. +\left( 2\lambda _{13}\nu _{\rho ^{3}}^{2}\nu _{\rho ^{2}}-\lambda
_{8}\nu _{\rho ^{3}}^{2}\nu _{\rho ^{2}}+2\lambda _{13}\nu _{\rho
^{2}}^{3}-\lambda _{8}\nu _{\rho ^{2}}^{3}\right) \nu _{\chi }\right] ,
\label{50}
\end{eqnarray}%
and their corresponding eigenvectors are

\begin{eqnarray}
V_{1} &\approx &\left( 
\begin{array}{c}
0 \\ 
0 \\ 
\nu _{\rho ^{2}}^{2} \\ 
\nu _{\eta }\nu _{\rho ^{3}} \\ 
\nu _{\chi }\nu _{\rho ^{2}}%
\end{array}%
\right) \;\;\;\;\;\;\;\;\;\;,\;\;\;\;\;\;\;\;V_{2}\approx \left( 
\begin{array}{c}
\nu _{\chi }\nu _{\rho ^{2}}^{2} \\ 
0 \\ 
0 \\ 
-\nu _{\eta }\left( \nu _{\rho ^{2}}^{2}+\nu _{\rho ^{3}}^{2}\right) \\ 
-\nu _{\chi }\nu _{\rho ^{2}}\nu _{\rho ^{3}}%
\end{array}%
\right) ,  \notag \\
V_{3} &\approx &\left( 
\begin{array}{c}
0 \\ 
\nu _{\rho ^{2}} \\ 
0 \\ 
-\nu _{\eta } \\ 
0%
\end{array}%
\right) \;\;\;\;\;\;\;\;\;\;\;,\;\;\;\;\;\;\;\;\quad V_{4}\approx \left( 
\begin{array}{c}
\nu _{\eta }\nu _{\rho ^{2}} \\ 
\nu _{\chi }\nu _{\eta } \\ 
0 \\ 
\nu _{\chi }\nu _{\rho ^{2}} \\ 
-\nu _{\rho ^{3}}\nu _{\eta }%
\end{array}%
\right) ,  \notag \\
V_{5} &\approx &\left( 
\begin{array}{c}
-\nu _{\rho ^{3}} \\ 
0 \\ 
\nu _{\chi } \\ 
0 \\ 
-\nu _{\rho ^{2}}%
\end{array}%
\right) .  \label{51}
\end{eqnarray}

\subsection{Diagonalization of the real sector\label{ap:diagre}}

The equation for the eigenvalues associated to the matrix (\ref{real}), has
no simple solutions. Then, we take the approximations (\ref{27}) and (\ref%
{28}), such that we keep only quadratic order in$\ \nu _{\chi }$ 
\begin{eqnarray}
&&\!\!\!\!\!\!\!\!%
\begin{array}{ccccc}
\qquad \mathbf{\xi }_{\chi } & \qquad \mathbf{\xi }_{\rho ^{2}}\qquad \qquad
& \qquad \mathbf{\xi }_{\rho ^{3}}\qquad \qquad & \qquad \mathbf{\xi }_{\eta
} & \qquad \mathbf{\xi }_{\chi ^{0}}%
\end{array}
\notag \\
M_{\xi \xi }^{2} &\approx &\left[ 
\begin{array}{ccccc}
8\lambda _{1}\nu _{\chi }^{2} & 0 & 4\lambda _{10}\nu _{\chi }^{2} & 0 & 0
\\ 
0 & -2f\frac{\nu _{\eta }\nu _{\chi }}{\nu _{\rho ^{2}}} & 0 & 2f\nu _{\chi }
& 0 \\ 
4\lambda _{10}\nu _{\chi }^{2} & 0 & \left( 2\lambda _{8}+4\lambda
_{13}\right) \nu _{\chi }^{2}-2f\frac{\nu _{\chi }\nu _{\eta }}{\nu _{\rho
^{2}}} & 0 & 0 \\ 
0 & 2f\nu _{\chi } & 0 & -2f\frac{\nu _{\chi }\nu _{\rho ^{2}}}{\nu _{\eta }}
& 0 \\ 
0 & 0 & 0 & 0 & 0%
\end{array}%
\right] ,  \label{53}
\end{eqnarray}

\noindent and the following decoupled submatrices arise

\begin{eqnarray}
&&%
\begin{array}{cc}
\qquad \mathbf{\xi }_{\rho ^{2}} & \qquad \mathbf{\xi }_{\eta }%
\end{array}
\notag \\
M_{\xi _{\rho ^{2}}\xi _{\eta }}^{2} &\approx &\left[ 
\begin{array}{cc}
-2f\frac{\nu _{\eta }\nu _{\chi }}{\nu _{\rho ^{2}}} & 2f\nu _{\chi } \\ 
2f\nu _{\chi } & -2f\frac{\nu _{\chi }\nu _{\rho ^{2}}}{\nu _{\eta }}%
\end{array}%
\right] ,  \notag  \label{16} \\
&&%
\begin{array}{ccccc}
\qquad \mathbf{\xi }_{\chi } & \qquad & \qquad \mathbf{\xi }_{\rho
^{3}}\qquad \qquad & \qquad & 
\end{array}
\notag \\
M_{\xi _{\chi }\xi _{\rho ^{3}}}^{2} &\approx &\left[ 
\begin{array}{cc}
8\lambda _{1}\nu _{\chi }^{2} & 4\lambda _{10}\nu _{\chi }^{2} \\ 
4\lambda _{10}\nu _{\chi }^{2} & \left( 2\lambda _{8}+4\lambda _{13}\right)
\nu _{\chi }^{2}-2f\frac{\nu _{\chi }\nu _{\eta }}{\nu _{\rho ^{2}}}%
\end{array}%
\right] ,  \notag \\
M_{\xi _{\chi ^{0}}}^{2} &=&0  \label{54}
\end{eqnarray}%
The first square matrix in Eq. (\ref{54}) has one null eigenvalue. However,
this first null eigenvalue arises due to the approximations given by Eqs. (%
\ref{27},\ref{28}) applied to (\ref{real}). Therefore, we take all terms in
the submatrix $M_{\xi _{\rho ^{2}}\xi _{\eta }}^{2}$ in Eq. (\ref{54}) 
\begin{eqnarray}
&&%
\begin{array}{cccccccccc}
&  &  & \quad \mathbf{\xi }_{\rho ^{2}} &  &  &  &  &  & \quad \mathbf{\xi }%
_{\eta }%
\end{array}
\notag \\
M_{\xi _{\rho ^{2}}\xi _{\eta }}^{2} &=&\left[ 
\begin{array}{cc}
8\lambda _{2}\nu _{\rho ^{2}}^{2}-2f\frac{\nu _{\eta }\nu _{\chi }}{\nu
_{\rho ^{2}}} & 4\lambda _{6}\nu _{\eta }\nu _{\rho ^{2}}+2f\nu _{\chi } \\ 
4\lambda _{6}\nu _{\eta }\nu _{\rho ^{2}}+2f\nu _{\chi } & 8\lambda _{3}\nu
_{\eta }^{2}-2f\frac{\nu _{\chi }\nu _{\rho ^{2}}}{\nu _{\eta }}%
\end{array}%
\right] ,  \label{55}
\end{eqnarray}%
whose eigenvalues are $P_{2}$ and $P_{3}$ in Eq. (\ref{36 a}). We should
observe precisely that $P_{2}$ is a small value that does not depend on $\nu
_{\chi }$. Owing to it, $P_{2}$ vanishes for the first approximation of $%
M_{\xi _{\rho ^{2}}\xi _{\eta }}^{2}$ in (\ref{54}).

\section{Bilinear Gauge-Goldstone bosons terms for $\protect\beta =1/\protect%
\sqrt{3}$\label{ap:acoplesGg}}

The kinetic term of the Higgs Lagrangian is: 
\begin{eqnarray}
\mathcal{L}_{\mathcal{H}} &=&\left( D^{\mu }\Phi \right) ^{\dagger }D_{\mu
}\Phi  \notag \\
&=&\overset{\mathbf{1}}{\overbrace{\left( \partial ^{\mu }\Phi \right)
^{\dagger }\left( D_{\mu }\Phi \right) +\left( D^{\mu }\Phi \right)
^{\dagger }\left( \partial _{\mu }\Phi \right) }}-\left( \partial ^{\mu
}\Phi \right) ^{\dagger }\left( \partial _{\mu }\Phi \right)  \notag \\
&&+\overset{\mathbf{2}}{\overbrace{\Phi ^{\dagger }\left( g\mathbf{W}^{\mu
}+g^{\prime }B^{\mu }X_{\Phi }\right) ^{\dagger }\left( g\mathbf{W}_{\mu
}+g^{\prime }B_{\mu }X_{\Phi }\right) \Phi },}  \label{B1}
\end{eqnarray}%
with 
\begin{equation}
D_{\mu }=\partial _{\mu }+igW_{\mu }^{\alpha }G_{\alpha }+ig^{\prime
}X_{\Phi }B_{\mu },  \label{B2}
\end{equation}%
The term \textbf{1} in Eq. (\ref{B1}) contains couplings among gauge fields
and the derivatives of the scalar fields $\partial _{\mu }\Phi ,$ from which
we obtain information about each would-be Goldstone boson coupled to the
corresponding massive gauge boson. The term \textbf{2} generates the masses
of the gauge bosons and couplings with the physical Higgs fields. The term 
\textbf{1} is evaluated to identify the would-be Goldstone bosons obtained
in the diagonalization of the mass matrix in the scalar sector, and its
corresponding gauge boson that acquires mass. The gauge bosons of the model
with $\beta =1/\sqrt{3}$ read 
\begin{eqnarray}
W_{\mu }^{\alpha }G_{\alpha } &=&\frac{1}{2}\left[ 
\begin{array}{ccc}
W_{\mu }^{3}+\frac{1}{\sqrt{3}}W_{\mu }^{8} & -\sqrt{2}W_{\mu }^{+} & \sqrt{2%
}K_{\mu }^{+} \\ 
-\sqrt{2}W_{\mu }^{-} & -W_{\mu }^{3}+\frac{1}{\sqrt{3}}W_{\mu }^{8} & \sqrt{%
2}K_{\mu }^{0} \\ 
\sqrt{2}K_{\mu }^{-} & \sqrt{2}\overline{K_{\mu }^{0}} & -\frac{2}{\sqrt{3}}%
W_{\mu }^{8}%
\end{array}%
\right] ,  \notag \\
&&\mathbf{B}_{\mu }=IB_{\mu }=\left[ 
\begin{array}{ccc}
B_{\mu } & 0 & 0 \\ 
0 & B_{\mu } & 0 \\ 
0 & 0 & B_{\mu }%
\end{array}%
\right] ,  \label{B3}
\end{eqnarray}%
where the neutral electroweak basis is related to the physical bosons by 
\begin{equation}
\left[ 
\begin{array}{c}
A_{\mu } \\ 
Z_{\mu } \\ 
Z_{\mu }^{\prime }%
\end{array}%
\right] =\left[ 
\begin{array}{ccc}
S_{W} & \frac{1}{\sqrt{3}}S_{W} & C_{W}\sqrt{1-\frac{1}{3}T_{W}^{2}} \\ 
-C_{W} & \frac{1}{\sqrt{3}}S_{W}T_{W} & S_{W}\sqrt{1-\frac{1}{3}T_{W}^{2}}
\\ 
0 & -\sqrt{1-\frac{1}{3}T_{W}^{2}} & \frac{1}{\sqrt{3}}T_{W}%
\end{array}%
\right] \left[ 
\begin{array}{c}
W_{\mu }^{3} \\ 
W_{\mu }^{8} \\ 
B_{\mu }%
\end{array}%
\right] ,  \label{B4}
\end{equation}%
where the Weinberg angle $\theta _{W}$ is defined by $\tan \theta _{W}=T_{W}=%
\frac{\sqrt{3}g^{\prime }}{\sqrt{3g^{2}+g^{\prime 2}}}.$

For the fields $\Phi =\chi $ and $\rho $ of table \ref{tab:diez} ($X_{\Phi }=%
\frac{1}{3}$), the covariant derivative Eq. (\ref{B2}) in the basis of the
physical gauge fields is

\begin{equation}
D_{\mu }=I\partial _{\mu }+i\left[ 
\begin{array}{ccc}
a_{1}A_{\mu }+a_{2}Z_{\mu }+a_{3}Z_{\mu }^{\prime } & -\frac{g}{\sqrt{2}}%
W_{\mu }^{+} & \frac{g}{\sqrt{2}}K_{\mu }^{+} \\ 
\frac{-g}{\sqrt{2}}W_{\mu }^{-} & a_{4}Z_{\mu }+a_{3}Z_{\mu }^{\prime } & 
\frac{g}{\sqrt{2}}K_{\mu }^{0} \\ 
\frac{g}{\sqrt{2}}K_{\mu }^{-} & \frac{g}{\sqrt{2}}\overline{K_{\mu }^{0}} & 
\frac{\sqrt{3g^{2}+g^{\prime 2}}}{3}Z_{\mu }^{\prime }%
\end{array}%
\right] ,  \label{B5}
\end{equation}%
For $\Phi =\eta $ ($X_{\Phi }=-\frac{2}{3}$), we have 
\begin{eqnarray}
D_{\mu } &=&I\partial _{\mu } \\
&+&i\left[ 
\begin{array}{ccc}
-a_{4}Z_{\mu }-a_{5}Z_{\mu }^{\prime } & \frac{-g}{\sqrt{2}}W_{\mu }^{+} & 
\frac{g}{\sqrt{2}}K_{\mu }^{+} \\ 
\frac{-g}{\sqrt{2}}W_{\mu }^{-} & -a_{1}A_{\mu }-a_{2}Z_{\mu }-a_{5}Z_{\mu
}^{\prime } & \frac{g}{\sqrt{2}}K_{\mu }^{0} \\ 
\frac{g}{\sqrt{2}}K_{\mu }^{-} & \frac{g}{\sqrt{2}}\overline{K_{\mu }^{0}} & 
-a_{4}A_{\mu }-a_{6}Z_{\mu }-2a_{3}Z_{\mu }^{\prime }%
\end{array}%
\right] ,  \notag
\end{eqnarray}%
and 
\begin{eqnarray}
a_1 &=&gS_W,\qquad a_2=\frac g2C_W\left( T_W^2-1\right) ,  \notag \\
a_3 &=&\frac{g^{\prime }}{2\sqrt{3}}\left( \frac{T_W^2-1}{T_W}\right)
,\qquad a_4=\frac g2C_W\left( T_W^2+1\right)  \notag \\
a_5 &=&\frac{g^{\prime }}{2\sqrt{3}}\left( \frac{T_W}{S_W^2}\right) ,\qquad
a_6=\frac{g^{\prime }}{\sqrt{3}}\left( \frac 1{T_W}\right) .  \label{B6}
\end{eqnarray}

With the covariant derivative for the fields $\chi $, $\rho $ and $\eta $,
we have that the terms of the bilinear scalar-gauge mixing coming from the
term \textbf{1} of Eq. (\ref{B1}) are given respectively by%
\begin{eqnarray}
&&\left( \partial ^{\mu }\chi \right) ^{\dagger }\left( D_{\mu }\left\langle
\chi \right\rangle _{0}\right) +h.c=  \label{bilinealx} \\
&&\frac{ig}{\sqrt{2}}\partial _{\mu }\left( \nu _{\chi }\chi ^{-}\right)
K_{\mu }^{+}-\frac{ig}{\sqrt{2}}\partial _{\mu }\left( \nu _{\chi }\chi
^{+}\right) K_{\mu }^{-}+\frac{ig}{\sqrt{2}}\partial _{\mu }\left( \nu
_{\chi }\xi _{\chi ^{0}}-i\nu _{\chi }\zeta _{\chi ^{0}}\right) K_{\mu }^{0}
\notag \\
&&-\frac{ig}{\sqrt{2}}\left( \nu _{\chi }\partial _{\mu }\xi _{\chi
^{0}}+i\nu _{\chi }\partial _{\mu }\zeta _{\chi ^{0}}\right) \overline{%
K_{\mu }^{0}}+\frac{\sqrt{3g^{2}+g^{\prime 2}}}{3}\partial _{\mu }\left(
2\nu _{\chi }\zeta _{\chi }\right) Z_{\mu }^{\prime },  \notag
\end{eqnarray}
\begin{eqnarray}
&&\left( \partial ^{\mu }\rho \right) ^{\dagger }\left( D_{\mu }\left\langle
\rho \right\rangle _{0}\right) +h.c=  \label{bilinealy} \\
&&\frac{ig}{\sqrt{2}}\partial _{\mu }\left( \nu _{\rho ^{3}}\rho ^{-}\right)
K_{\mu }^{+}-\frac{ig}{\sqrt{2}}\left( \partial _{\mu }\nu _{\rho ^{3}}\rho
^{+}\right) K_{\mu }^{-}-\frac{ig}{\sqrt{2}}\partial _{\mu }\left( \nu
_{\rho ^{2}}\rho ^{-}\right) W_{\mu }^{+}  \notag \\
&&+\frac{ig}{\sqrt{2}}\partial _{\mu }\left( \nu _{\rho ^{2}}\rho
^{+}\right) W_{\mu }^{-}+\frac{ig}{\sqrt{2}}\partial _{\mu }\left( \nu
_{\rho ^{3}}\xi _{\rho ^{2}}-\nu _{\rho ^{2}}\xi _{\rho ^{3}}-i\nu _{\rho
^{3}}\zeta _{\rho ^{2}}-i\nu _{\rho ^{2}}\zeta _{\rho ^{3}}\right) K_{\mu
}^{0}  \notag \\
&&-\frac{ig}{\sqrt{2}}\partial _{\mu }\left( \nu _{\rho ^{3}}\xi _{\rho
^{2}}-\nu _{\rho ^{2}}\xi _{\rho ^{3}}+i\nu _{\rho ^{3}}\zeta _{\rho
^{2}}+i\nu _{\rho ^{2}}\zeta _{\rho ^{3}}\right) \overline{K_{\mu }^{0}}%
+a_{4}\partial _{\mu }\left( 2\nu _{\rho ^{2}}\zeta _{\rho ^{2}}\right)
Z_{\mu }  \notag \\
&&+\frac{\sqrt{3g^{2}+g^{\prime 2}}}{3}\partial _{\mu }\left( 2\nu _{\rho
^{3}}\zeta _{\rho ^{3}}+\frac{6a_{3}}{\sqrt{3g^{2}+g^{\prime 2}}}\nu _{\rho
^{2}}\zeta _{\rho ^{2}}\right) Z_{\mu }^{\prime },  \notag
\end{eqnarray}%
\begin{eqnarray}
&&\left( \partial ^{\mu }\eta \right) ^{\dagger }\left( D_{\mu }\left\langle
\eta \right\rangle _{0}\right) +h.c=  \label{bilinealz} \\
&&-\frac{ig}{\sqrt{2}}\partial _{\mu }\left( \nu _{\eta }\eta ^{3-}\right)
K_{\mu }^{+}+\frac{ig}{\sqrt{2}}\partial _{\mu }\left( \nu _{\eta }\eta
^{3+}\right) K_{\mu }^{-}+\frac{ig}{\sqrt{2}}\partial _{\mu }\left( \nu
_{\eta }\eta ^{2-}\right) W_{\mu }^{+}  \notag \\
&&-\frac{ig}{\sqrt{2}}\partial _{\mu }\left( \nu _{\eta }\eta ^{2+}\right)
W_{\mu }^{-}-a_{4}\partial _{\mu }\left( 2\nu _{\eta }\zeta _{\eta }\right)
Z_{\mu }-\frac{a_{4}\sqrt{3g^{2}+4g^{\prime 2}}}{3g}\partial _{\mu }\left(
2\nu _{\eta }\zeta _{\eta }\right) Z_{\mu }^{\prime }.  \notag
\end{eqnarray}%
Adding the Eqs. (\ref{bilinealx}-\ref{bilinealz}) we can find the would-be
Goldstone bosons in the $R_{\zeta }$ gauge associated to the gauge fields
that acquire masses 
\begin{eqnarray}
&&\left( \partial ^{\mu }\Phi \right) ^{\dagger }\left( D_{\mu }\left\langle
\Phi \right\rangle _{0}\right) +h.c=  \notag \\
&&\frac{ig}{\sqrt{2}}\partial _{\mu }\left( -\nu _{\rho ^{2}}\rho ^{-}+\nu
_{\eta }\eta ^{2-}\right) W_{\mu }^{+}-\frac{ig}{\sqrt{2}}\partial _{\mu
}\left( -\nu _{\rho ^{2}}\rho ^{+}+\nu _{\eta }\eta ^{2+}\right) W_{\mu }^{-}
\notag \\
&+&\frac{ig}{\sqrt{2}}\partial _{\mu }\left( \nu _{\chi }\chi ^{-}+\nu
_{\rho ^{3}}\rho ^{-}-\nu _{\eta }\eta ^{3-}\right) K_{\mu }^{+}-\frac{ig}{%
\sqrt{2}}\partial _{\mu }\left( \nu _{\chi }\chi ^{+}+\nu _{\rho ^{3}}\rho
^{+}-\nu _{\eta }\eta ^{3+}\right) K_{\mu }^{-}  \notag \\
&+&\frac{ig}{\sqrt{2}}\partial _{\mu }\left( \nu _{\chi }\xi _{\chi
^{0}}+\nu _{\rho ^{3}}\xi _{\rho ^{2}}-\nu _{\rho ^{2}}\xi _{\rho ^{3}}-i\nu
_{\chi }\zeta _{\chi ^{0}}-i\nu _{\rho ^{3}}\zeta _{\rho ^{2}}-i\nu _{\rho
^{2}}\zeta _{\rho ^{3}}\right) K_{\mu }^{0}  \notag \\
&-&\frac{ig}{\sqrt{2}}\partial _{\mu }\left( \nu _{\chi }\xi _{\chi
^{0}}+\nu _{\rho ^{3}}\xi _{\rho ^{2}}-\nu _{\rho ^{2}}\xi _{\rho ^{3}}+i\nu
_{\chi }\zeta _{\chi ^{0}}+i\nu _{\rho ^{3}}\zeta _{\rho ^{2}}+i\nu _{\rho
^{2}}\zeta _{\rho ^{3}}\right) \overline{K_{\mu }^{0}}  \notag \\
&+&\frac{g\sqrt{3g^{2}+4g^{\prime 2}}}{\sqrt{3g^{2}+g^{\prime 2}}}\partial
_{\mu }\left( \nu _{\rho ^{2}}\zeta _{\rho ^{2}}-\nu _{\eta }\zeta _{\eta
}\right) Z_{\mu }  \label{kinx} \\
&+&\frac{2\sqrt{3g^{2}+g^{\prime 2}}}{3}\partial _{\mu }\left( \nu _{\chi
}\zeta _{\chi }+\nu _{\rho ^{3}}\zeta _{\rho ^{3}}+\frac{1}{2}\frac{%
2g^{\prime 2}-3g^{2}}{3g^{2}+g^{\prime 2}}\nu _{\rho ^{2}}\zeta _{\rho ^{2}}-%
\frac{1}{2}\frac{3g^{2}+4g^{\prime 2}}{3g^{2}+g^{\prime 2}}\nu _{\eta }\zeta
_{\eta }\right) Z_{\mu }^{\prime }.  \notag
\end{eqnarray}

In the couplings with $K_{\mu }^{0}$ and $\overline{K_{\mu }^{0}}$, it is
observed a mixing amongst the real and imaginary scalars, which is not
produced in the mass matrices of the scalar sector, as it is confirmed by
Eqs. (\ref{imaginario}) and (\ref{real}). To have it, we make the rotation 
\begin{eqnarray}
&&\frac{ig}{\sqrt{2}}\partial _{\mu }\left( \nu _{\chi }\xi _{\chi
^{0}}-i\nu _{\chi }\zeta _{\chi ^{0}}+\nu _{\rho ^{3}}\xi _{\rho ^{2}}-\nu
_{\rho ^{2}}\xi _{\rho ^{3}}-i\nu _{\rho ^{3}}\zeta _{\rho ^{2}}-i\nu _{\rho
^{2}}\zeta _{\rho ^{3}}\right) K_{\mu }^{0}  \notag \\
&&-\frac{ig}{\sqrt{2}}\partial _{\mu }\left( \nu _{\chi }\xi _{\chi
^{0}}+i\nu _{\chi }\zeta _{\chi ^{0}}+\nu _{\rho ^{3}}\xi _{\rho ^{2}}-\nu
_{\rho ^{2}}\xi _{\rho ^{3}}+i\nu _{\rho ^{3}}\zeta _{\rho ^{2}}+i\nu _{\rho
^{2}}\zeta _{\rho ^{3}}\right) \overline{K_{\mu }^{0}}  \notag \\
&=&\frac{ig}{\sqrt{2}}\partial _{\mu }\left( \nu _{\chi }\xi _{\chi
^{0}}+\nu _{\rho ^{3}}\xi _{\rho ^{2}}-\nu _{\rho ^{2}}\xi _{\rho
^{3}}\right) \widetilde{K_{\mu }^{0}}  \label{kiny} \\
&&-\frac{ig}{\sqrt{2}}\partial _{\mu }\left( i\nu _{\chi }\zeta _{\chi
^{0}}+i\nu _{\rho ^{3}}\zeta _{\rho ^{2}}+i\nu _{\rho ^{2}}\zeta _{\rho
^{3}}\right) \overline{\widetilde{K_{\mu }^{0}}}.  \notag
\end{eqnarray}%
with $\widetilde{K_{\mu }^{0}}=K_{\mu }^{0}-\overline{K_{\mu }^{0}}$ and $%
\overline{\widetilde{K_{\mu }^{0}}}=K_{\mu }^{0}+\overline{K_{\mu }^{0}}.$
The couplings in Eqs. (\ref{kinx}) and (\ref{kiny}), indicates the following
combinations of the would-be Goldstone bosons

\begin{eqnarray}
\phi _{1}^{\pm } &\sim &\nu _{\rho ^{2}}\rho ^{\pm }-\nu _{\eta }\eta ^{2\pm
}  \notag \\
\phi _{2}^{\pm } &\sim &\nu _{\chi }\chi ^{\pm }+\nu _{\rho ^{3}}\rho ^{\pm
}-\nu _{\eta }\eta ^{3\pm }  \notag \\
\phi _{1}^{0} &\sim &\left( \nu _{\chi }\zeta _{\chi ^{0}}+\nu _{\rho
^{3}}\zeta _{\rho ^{2}}+\nu _{\rho ^{2}}\zeta _{\rho ^{3}}\right)  \notag \\
\phi _{2}^{0} &\sim &\left( \nu _{\chi }\zeta _{\chi }+\nu _{\rho ^{3}}\zeta
_{\rho ^{3}}+\frac{1}{2}\frac{2g^{\prime 2}-3g^{2}}{3g^{2}+g^{\prime 2}}\nu
_{\rho ^{2}}\zeta _{\rho ^{2}}-\frac{1}{2}\frac{3g^{2}+4g^{\prime 2}}{%
3g^{2}+g^{\prime 2}}\nu _{\eta }\zeta _{\eta }\right)  \notag \\
\phi _{3}^{0} &\sim &\left( \nu _{\rho ^{2}}\zeta _{\rho ^{2}}-\nu _{\eta
}\zeta _{\eta }\right)  \notag \\
\phi _{4}^{0} &\sim &\nu _{\chi }\xi _{\chi ^{0}}+\nu _{\rho ^{3}}\xi _{\rho
^{2}}-\nu _{\rho ^{2}}\xi _{\rho ^{3}}  \label{B9}
\end{eqnarray}%
that couples to the corresponding physical gauge bosons$\ W_{\mu }^{\pm }$,\ 
$K_{\mu }^{\pm }$,$\ \overline{\widetilde{K_{\mu }^{0}}\text{,\ }}Z_{\mu
}^{\prime }$,$\ Z_{\mu }$,$\ \widetilde{K_{\mu }^{0}}$ respectively.

Rotating the VEV to the basis where $\nu _{\rho ^{3}}\rightarrow 0$, as in
Eq. (\ref{rotVEV}), it is found

\begin{eqnarray}
\phi _{1}^{\pm } &\sim &\nu _{\rho ^{2}}\rho ^{\pm }-\nu _{\eta }\eta ^{2\pm
}  \notag \\
\phi _{2}^{\pm } &\sim &\nu _{\chi }\chi ^{\pm }-\nu _{\eta }\eta ^{3\pm } 
\notag \\
\phi _{1}^{0} &\sim &i\left( \nu _{\chi }\zeta _{\chi ^{0}}+\nu _{\rho
^{2}}\zeta _{\rho ^{3}}\right)  \notag \\
\phi _{2}^{0} &\sim &i\left( \nu _{\chi }\zeta _{\chi }+\frac{1}{2}\frac{%
2g^{\prime 2}-3g^{2}}{3g^{2}+g^{\prime 2}}\nu _{\rho ^{2}}\zeta _{\rho ^{2}}-%
\frac{1}{2}\frac{3g^{2}+4g^{\prime 2}}{3g^{2}+g^{\prime 2}}\nu _{\eta }\zeta
_{\eta }\right)  \notag \\
\phi _{3}^{0} &\sim &i\left( \nu _{\rho ^{2}}\zeta _{\rho ^{2}}-\nu _{\eta
}\zeta _{\eta }\right)  \notag \\
\phi _{4}^{0} &\sim &\nu _{\chi }\xi _{\chi ^{0}}-\nu _{\rho ^{2}}\xi _{\rho
^{3}}  \label{B10}
\end{eqnarray}

\end{document}